\documentclass[final,1p,times]{elsarticle}

\usepackage{amssymb,amsmath,bm}

\usepackage{tabularx}
\usepackage{nicefrac}
\usepackage{xspace}
\usepackage{xifthen}
\usepackage{fancyvrb}

\DefineVerbatimEnvironment{code}{Verbatim}{fontsize=\footnotesize}

\newcommand{\nanobem}[0]{\textsc{nanobem}\xspace}
\newcommand{\bem}[0]{\textsc{bem}\xspace}
\newcommand{\cim}[0]{\textsc{cim}\xspace}

\newcommand{\mmatrix}[2][]{
  \ifthenelse{\isempty{#1}}
    {\left\llbracket{#2}\right\rrbracket}
    {#1\llbracket{#2}#1\rrbracket}
}
\newcommand{\dy}[1]{\overset\leftrightarrow{\bm #1}} 

\newcounter{bla}

\journal{Computer Physics Communications}

\begin{document}

\begin{frontmatter}



\title{Nanophotonic resonance modes with the \nanobem toolbox}


\author[a]{Ulrich Hohenester\corref{author}}
\author[a]{Nikita Reichelt}
\author[a,b]{Gerhard Unger}

\cortext[author] {Corresponding author.\\\textit{E-mail address:} ulrich.hohenester@uni-graz.at}
\address[a]{Institute of Physics, University of Graz, Universit\"atsplatz 5, 8010 Graz, Austria}
\address[b]{Institute of Advanced Mathemathics, Technical University of Graz, Steyrergasse 30, 8010 Graz, Austria}

\begin{abstract}
\nanobem is a \textsc{matlab} toolbox for the solution of Maxwell's equations for nanophotonic systems and the computation of resonance modes, sometimes also referred to as quasinormal modes or resonance states.  It is based on a Galerkin scheme for the boundary element method, using Raviart-Thomas shape elements for the representation of the tangential electromagnetic fields at the particle boundary.  The toolbox is written in an object-oriented manner with the focus on clarity rather than speed, and has been developed and tested for small to intermediate problems with a few thousand boundary elements.  The computation of the resonance modes uses the contour integral method of Beyn.
\end{abstract}


\end{frontmatter}



{\bf PROGRAM SUMMARY}

\begin{small}
\noindent
{\em Program Title:}  \nanobem \\
{\em CPC Library link to program files:} (to be added by Technical Editor) \\
{\em Developer's repository link:} (if available) \\
{\em Code Ocean capsule:} (to be added by Technical Editor)\\
{\em Licensing provisions(please choose one):} GNU General Public License 3  \\
{\em Programming language:}   Matlab                 \\
{\em Supplementary material:}                                 \\
{\em Nature of problem:}
Solve Maxwell’s equations and compute resonance modes for optical resonators and nanophotonic systems with linear, homogeneous, and local materials separated by abrupt interfaces. \\
{\em Solution method:} Galerkin implementation of boundary element method approach using Raviart-Thomas shape elements, and contour integral method for the computation of nanophotonic resonance modes. \\
{\em Additional comments including restrictions and unusual features:}
Toolbox has been developed and tested for small to intermediate problems with a few thousand boundary elements.\\
\end{small}


\section{Introduction}\label{sec:intro}

Resonance modes, sometimes also referred to as quasinormal modes or resonant states, have recently received considerable interest in the field of nanophotonics~\cite{leung:94,kristensen:12,bai:13,sauvan:13,ge:14,doost:14,muljarov:16,perrin:16,lalanne:19,kristensen:20}.  In optical resonators and cavities, light can be confined on a length scale comparable to the wavelength.  By binding light to coherent excitations of nanostructures, such as surface plasmons or surface phonons, one can achieve light confinement down to extreme sub-wavelength volumes~\cite{schuller:10,hohenester:20}.   Such optical resonators play an important role in science and engineering, for instance for microwave resonators, semiconductor lasers, or optical microcavities and plasmonic nano-cavities.

From a computational perspective, the simulation of optical resonators builds on the solution of Maxwell's equations.  Let $u$ denote the degrees of freedom for the representation of the (total or scattered) electromagnetic fields, e.g. curl-conforming Nedelec elements for the finite element method approach~\cite{hesthaven:02,hesthaven:03} or Raviart-Thomas shape elements for the boundary element method approach~\cite{chew:95,kern:09,hohenester:20}, and $q_{\rm inc}$ the incoming electromagnetic fields, such as a planewave excitation of the optical resonator.  For the linear material properties of our present concern, $u$ is related to an excitation $q_{\rm inc}$ oscillating with angular frequency $\omega$ through
\begin{equation}\label{eq:intro1}
  A(\omega)u=q_{\rm inc}\,,
\end{equation}
where the transmission matrix $A(\omega)$ has to be derived from Maxwell's equations subject to the proper boundary conditions between different materials.  Solving Eq.~\eqref{eq:intro1} for the unknowns $u$ provides us with the solutions of Maxwell's equations.  Instead of solving Eq.~\eqref{eq:intro1} directly, we can also seek for the resonance frequencies $\omega_k$ and modes $v_k$ of the transmission problem, which are defined such that they can persist in absence of an external excitation,
\begin{equation}\label{eq:intro2}
  A(\omega_k)v_k=0\,.
\end{equation}
For absorbing media or open cavities, Eq.~\eqref{eq:intro2} can only be fulfilled for complex frequencies $\omega_k=\omega_k'+i\omega_k''$, where the material and optical losses are compensated by the imaginary part $\omega_k''$.  In a second step, the electromagnetic response $u$ to the incoming fields $q_{\rm inc}$ is expressed in terms of these resonance modes
\begin{equation}\label{eq:intro3}
  u\approx\sum_k C_k(\omega)v_k\,,
\end{equation}

\begin{figure}[t]
\includegraphics[width=\textwidth]{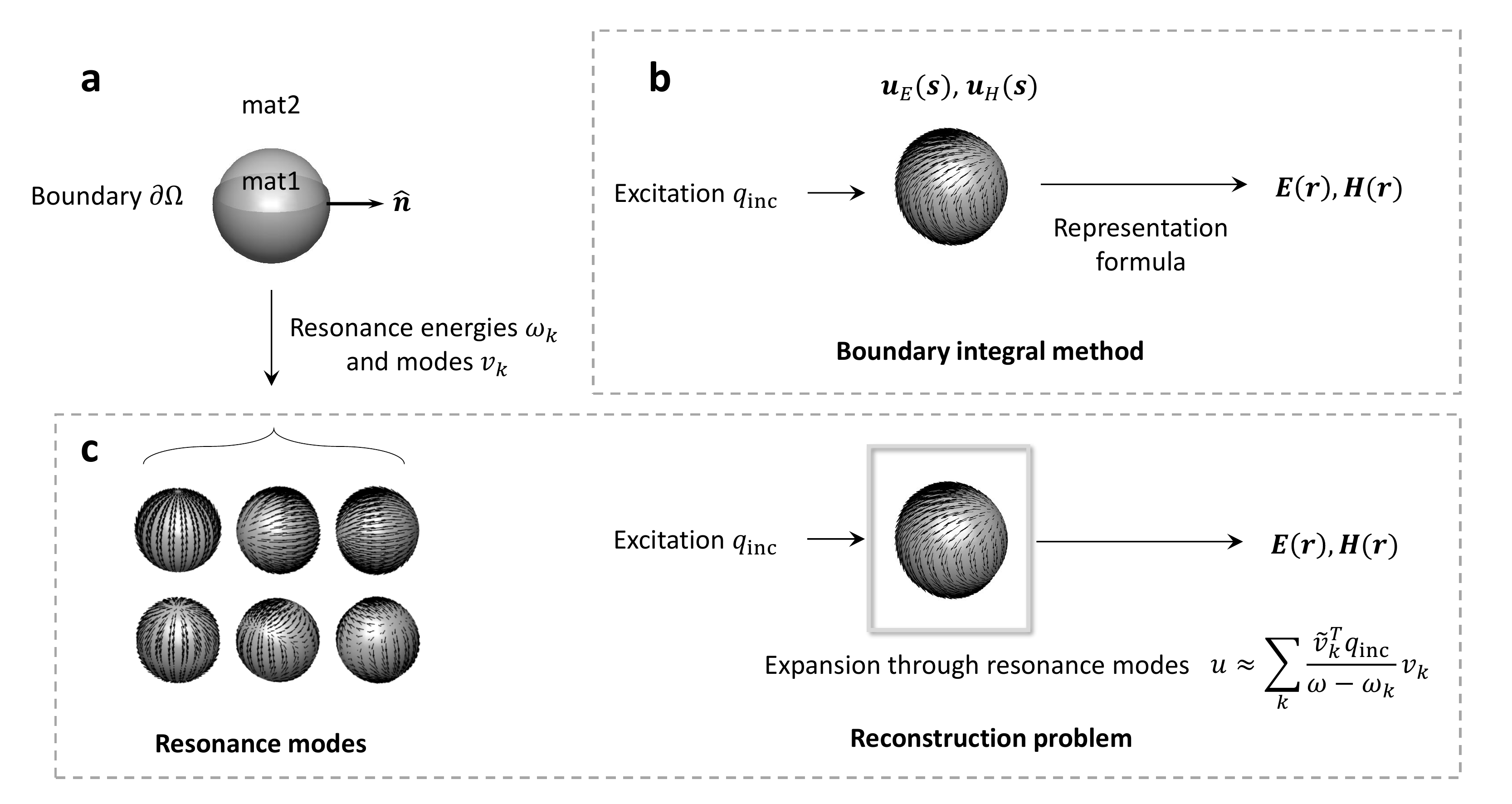}
\caption{Schematics of the boundary integral method (\textsc{bim}) and the resonance modes.  (a) We consider a setup where a nanoparticle with homogeneous material parameters \texttt{mat1} is embedded in a medium with material parameters \texttt{mat2}.  The boundary separating the two media is denoted with $\partial\Omega$, the outer surface normal of the boundary is $\hat{\bm n}$.  (b) In \textsc{bim}, the response of the nanoparticle to some external perturbation (e.g. planewave excitation or oscillating dipole) can be described in terms of the tangential electromagnetic fields $\bm u_{E,H}$ at the particle boundary, fields away from the boundary can be computed from the representation formula.  (c) In the resonance mode approach we don't solve Maxwell's equation directly, but first determine the resonance frequencies $\omega_k$ and modes $v_k$ of the nanoparticle, and then expand the tangential fields $\bm u_{E,H}$ in terms of these modes.  The second step is usually referred to as the ``reconstruction problem''.}
\label{fig:bemschem1}
\end{figure}

\noindent where the determination of the expansion coefficients $C_k(\omega)$ is usually referred to as the reconstruction problem~\cite{lalanne:19}, see also Fig.~\ref{fig:bemschem1} for a schematic representation.  The advantage of the resonance mode solution~\eqref{eq:intro3} is often the speedup in comparison to the direct solution of Eq.~\eqref{eq:intro1}, provided that the resonance modes are at hand.  Probably even more important is the interpretation of the solutions in terms of intuitive and physically transparent resonance modes, which also govern the physics of optical resonators.  These modes can be additionally submitted to a quantization procedure~\cite{franke:19}, which is beneficial for quantum optical investigations.

The consideration of complex frequencies in Eq.~\eqref{eq:intro2} for the evaluation of the resonance modes can be complicated for volume-based schemes, such as the finite element method approach, because of the increase of the electromagnetic fields away from the resonators.  Also the damping of such fields in perfectly matched layers surrounding the simulation volume can be critical.  Although these problems can be overcome in principle, approaches such as the boundary element method (\bem) avoid these difficulties because the solution vector $u$ only involves the tangential electromagnetic fields at the boundaries separating different materials.  As has been demonstrated in~\cite{unger:18}, the determination of resonance modes using a \bem approach is relatively simple and free from any conceptional or computational difficulties.

In this paper we present the \nanobem toolbox for the computation of resonance modes using a \bem approach based on a Galerkin scheme~\cite{chew:95,kern:09,hohenester:20}.  The toolbox is implemented in \textsc{matlab} in an object-oriented manner, and comes along with a number of demo files together with detailed help pages.  The main purpose of the toolbox is to provide a fast, efficient, and flexible toolkit for computing and analyzing resonance modes for simple nanophotonic structures, such as dielectric or metallic nanospheres or nanorods, and to allow for easy modifications and extensions in case of future improvements.  At the same time, the toolbox supersedes the \textsc{mnpbem} toolbox developed by one of the authors of this paper~\cite{hohenester.cpc:12,hohenester.cpc:14b,hohenester.cpc:18}, which is based on a collocation approach for the electrodynamic potentials~\cite{garcia:02}.  The user interface and the program structures are similar for the two toolboxes, but all internal classes have been completely rewritten for the \nanobem toolbox to accommodate for the Galerkin approach using the electromagnetic fields as basic quantities. 

We have organized this work as follows.  In Sec.~\ref{sec:start} we show how to set up the toolbox and how to run simple simulations for nanophotonic systems.  The theory underlying the \bem approach and the contour integral method (\cim) for computing the resonance modes is presented in Sec.~\ref{sec:theory}, implementation details of the toolbox are discussed in Sec.~\ref{sec:toolbox}.  Finally, in Sec.~\ref{sec:cimresults} we list a few dos and don'ts for the \cim approach and the computation of resonance modes.  Some theoretical and computational details are postponed to the Appendices.

\section{Getting started}\label{sec:start}

In this section, we briefly describe how to set up the toolbox, and how to run simple boundary element method (\bem) simulations and simulations using resonance modes.  To install the toolbox, one must add the path of the main directory \texttt{nanobemdir} of the \nanobem toolbox as well as the paths of all subdirectories to the \textsc{matlab} search path.  This can be done, for instance, through
\begin{code}
>> addpath(genpath(nanobemdir));
\end{code}
To set up the help pages, one must once change to the main directory of the \nanobem toolbox and run \texttt{makehelp}
\begin{code}
>> cd nanobemdir;
>> makehelp;
\end{code}
Once this is done, the help pages, which provide detailed information about the toolbox, are available in the \textsc{matlab} help browser.

\subsection{A simple BEM example}\label{sec:startbem}

Figure~\ref{fig:bemschem1}(a) shows a typical \bem example, where a gold nanosphere with a diameter of 50 nm is embedded in a dielectric medium representative for water, with a refractive index of $1.33$.  In the following we briefly discuss a simulation where the optical cross section is computed.  The corresponding \textsc{matlab} files are available as \verb!demoretspec01.m! and \verb!democim01.m! in the help pages of the toolbox.  The nanophotonic environment is set up through
\begin{code}
mat1 = Material( 1.33 ^ 2, 1 );          %  material properties for water
mat2 = Material( epsdrude( 'Au' ), 1 );  %  material properties for gold
mat = [ mat1, mat2 ];                    %  unique material vector
diameter = 50;                           %  diameter of sphere in nanometers
p = trisphere( 144, diameter );          %  triangulated nanosphere boundary
tau = BoundaryEdge( mat, p, [ 2, 1 ] );  %  boundary elements including shape elements
\end{code}
In the first two lines we set up material objects \verb!Material(eps,mu)!, where \verb!eps! is either a permittivity function or constant and \verb!mu! a permeability function or constant.  The \verb!epsdrude! object provides a Drude dielectric function representative for gold.  We finally specify a material vector \verb!mat! for the nanophotonic environment, where the first entry specifies the material properties of the background medium, and the other entries specify the material properties of the particles.  From here on, the same material vector should be passed to all toolbox functions.  See also Sec.~\ref{sec:bem}, \ref{sec:details}, and the help pages for a more detailed description.  We next generate a triangulation of the nanosphere boundary with 144 vertices.  With
\begin{code}
>> plot( p, 'EdgeColor', 'b' );
>> plot( p, 'EdgeColor', 'b', 'nvec', 1 );  %  show outer surface normals
\end{code}
we can plot the triangulated boundary and the outer surface normals.  Finally, we create a \verb!BoundaryEdge! object with the material vector, the triangulated boundary, and the information that the material at the boundary inside (measured with respect to the outer surface normal) is \verb!mat(2)! and the material at the boundary outside is \verb!mat(1)!.  The boundary elements \verb!tau! are additionally equipped with shape elements for the representation of the tangential electromagnetic fields, see also Fig.~\ref{fig:bemschem02}, as will be discussed in more detail below.

\begin{figure}
\includegraphics[width=\textwidth]{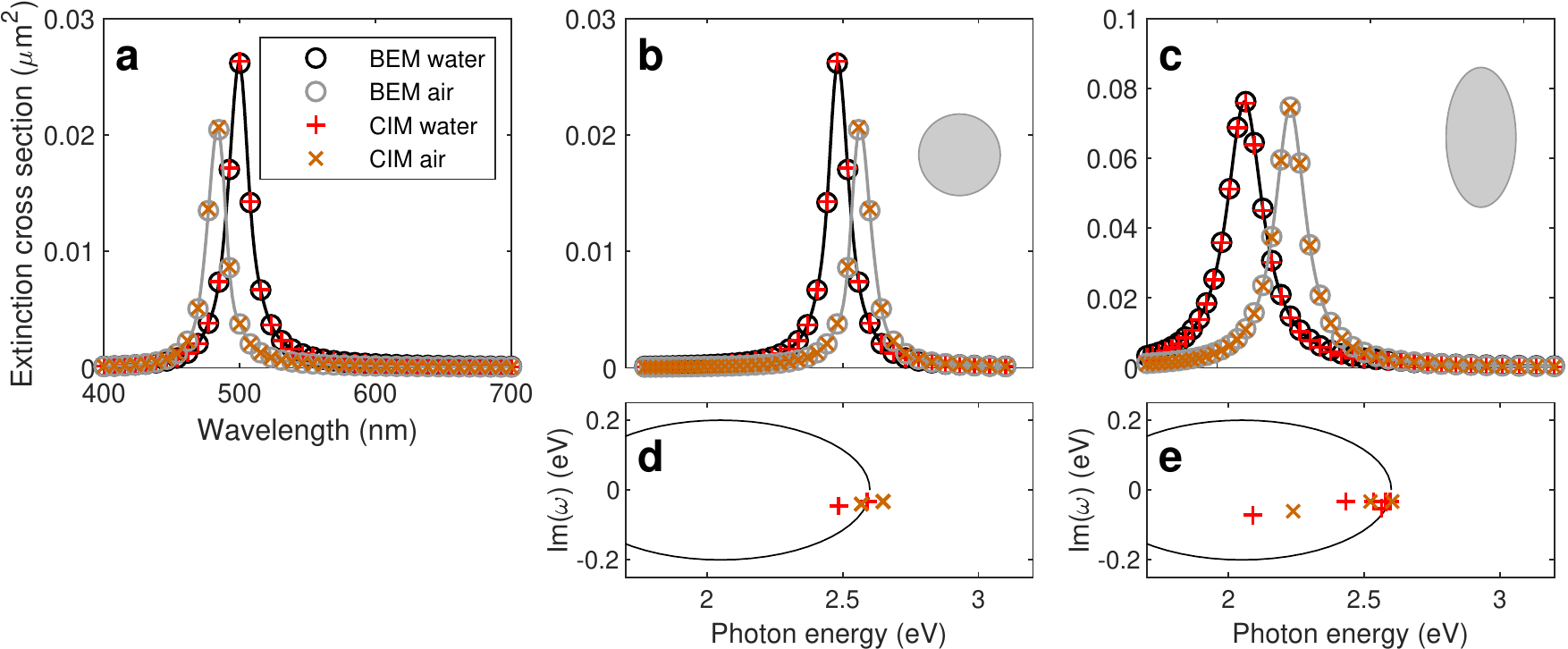}
\caption{(a) Extinction cross sections for gold nanosphere (diameter 50 nm) embedded in water (black line) and air (gray line), as obtained from the solutions of the \bem equations (circles) and the resonance mode expansion ($+$, $\times$).  (b) Same as (a) but using for the abscissa the photon energy in eV rather than the light wavelength in nanometers.  (c) Extinction cross sections for a gold nanoellipsoid with a long axis of 100 nm and short axis of 50 nm.  (d,e) \cim contour in the complex frequency plane (solid lines) and resonance mode energies $\omega_k$ (symbols).  Note that for small nanoparticles only modes with a non-vanishing dipole moment can be excited optically, with a similar behavior for larger nanoparticles where typically only a few modes contribute noticeably to the optical spectra.}
\label{fig:cim01}
\end{figure}

\begin{figure}[t]
\includegraphics[width=\textwidth]{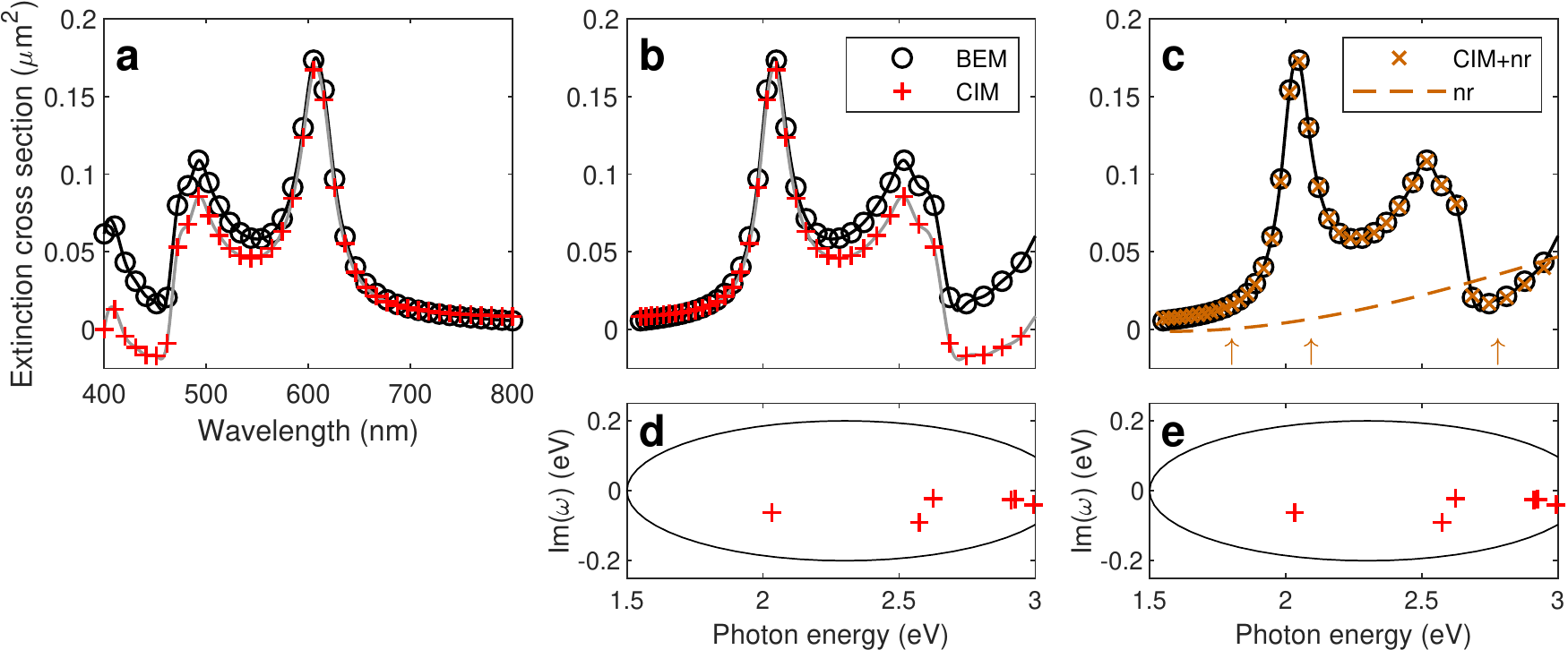}
\caption{Same as Fig.~\ref{fig:cim01} but for a dielectric nanosphere with a diameter of 150 nm.  For the nanosphere we consider a dielectric function representative for silicon, with material parameters taken from Ref.~\cite{sauvan:21}, which is embedded in a background medium with a refractive index of one.  (a,b) The \bem and \cim results using the 42 resonance modes shown in panel (d) agree well for smaller photon energies, say below 2.2 eV, but start to deviate consistently for larger energies.  (c,e) By considering in \cim the same resonance modes as in panel (d) together with a non-resonant background (dashed line), one obtains almost perfect agreement for all photon energies.  For details see discussion in Sec.~\ref{sec:cimsolver}.}
\label{fig:cim02}
\end{figure}

Once the nanophotonic environment is set up, we can run a simulation by specifying the external excitation, for instance a planewave excitation, initializing the \bem solver, and solving the \bem equations for a number of light wavelengths $\lambda$.  In the toolbox we use the free-space wavenumber $k_0=\nicefrac{2\pi}\lambda$ instead.
\begin{code}
pol = [ 1, 0, 0 ];                     %  light polarization
dir = [ 0, 0, 1 ];                     %  light propagation direction
exc = galerkin.planewave( pol, dir );  %  planewave excitation
bem = galerkin.bemsolver( tau );       %  BEM solver
lambda = linspace( 400, 800, 20 );
k0 = 2 * pi ./ lambda;                 %  wavenumber of light in vacuum
cext = zeros( size( k0 ) );            %  initialize array for extinction cross section

for i = 1 : numel( k0 )                %  loop over wavenumbers
  sol = bem \ exc( tau, k0( i ) );     %  solve BEM equations
  cext( i ) = extinction( exc, sol );  %  compute exctinction cross section
end
\end{code}
The above code is quite self-explanatory and we will discuss the various commands in more detail in Sec.~\ref{sec:bem}.  When solving the BEM equations, we obtain a solution object \verb!sol! that stores the tangential electromagnetic fields at the sphere boundary.  From these fields all other electromagnetic quantities can be determined, such as the extinction cross section.  We can finally plot the extinction cross section
\begin{code}
>> plot( lambda, cext );
\end{code}
to obtain the extinction cross section shown in Fig.~\ref{fig:cim01}(a).  See also Fig.~\ref{fig:cim02} for the example of a dielectric nanosphere with a Drude-Lorentz dielectric function representative for silicon.

\subsection{Resonance modes and the \cim solver}\label{sec:startcim}

We next show how to compute the optical cross sections using the resonance mode expansion. Details of our theoretical approach and of the working principle for the contour integral method (\cim) approach will be presented in Secs.~\ref{sec:theory} and \ref{sec:toolbox}, respectively.  In brief, we are seeking for a description of the nanophotonic environment in terms of (complex) resonance frequencies $\omega_k$ and modes, as schematically shown in Fig.~\ref{fig:bemschem1}(c).  To find these resonances, the user must define a contour in the complex frequency space, see ellipse in Figs.~\ref{fig:cim01}(d,e), and the \cim algorithm then returns the resonances located inside this region in an (almost) black-box manner.  We will discuss later in Sec.~\ref{sec:cimresults} that the approach only works for permittivity and permeability functions that can be extended to the complex frequency space, such as the Drude dielectric function used in the example of Sec.~\ref{sec:startbem}, and the contour must not cross branch cuts of the square root of these functions, and must also exclude accumulation points of resonances.  Within the toolbox, the \cim solver is set up with
\begin{code}
cim = galerkin.cimsolver( tau );
cim.contour = cimbase.ellipse( [ 1.5, 2.6 ], 0.2, 60 );
\end{code}

\begin{figure}
\includegraphics[width=\textwidth]{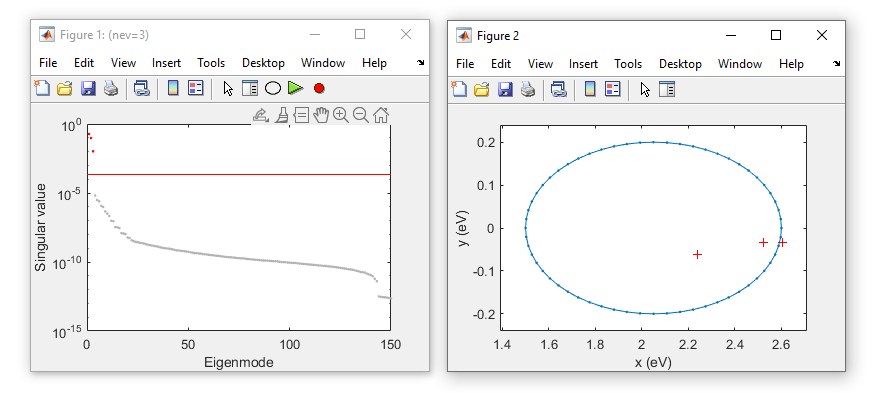}
\caption{Graphical user interface of the \texttt{tolselect} toolbox function for the computation of the resonance modes using the approach of Beyn~\cite{beyn:12}.  The left panel shows the singular values of the Beyn approach, see Sec.~\ref{sec:cimsolver} for details, and the user can interactively select those resonance modes that have a sufficiently large singular value.  This step is somewhat critical, because several things can go wrong in the \cim approach, as briefly discussed in Sec.~\ref{sec:cimresults}.  In the panel of the left \textsc{matlab} figure three special icons have been added: when the ellipse is pressed, the figure shown in the right panel opens up and displays the contour in the complex plane together with the resonance frequencies of the modes selected by the user.  When the green triangle $\triangleright$ is pressed, the \cim calculation continues, when the red circle $\bullet$ is pressed, the \cim calculation stops.}\label{fig:cimgui}
\end{figure}

\noindent In the second line we define an ellipse in the complex frequency plane $\omega=\omega'+i\omega''$, where the axis along $\omega'$ goes from 1.5 eV to 2.6 eV, and the axis for $\omega''$ goes from $-0.2$ eV to $0.2$ eV.  For the evaluation of the resonance modes we use 60 quadrature points, following the approach of Beyn~\cite{beyn:12} to be discussed in Sec.~\ref{sec:cimsolver}.  We then solve the \bem equations in the complex frequency plane, and the user must select the resonance modes to be kept in the simulations from a singular value decomposition~\cite{beyn:12}.  This is accomplished with
\begin{code}
data = eval1( cim );                        %  solve BEM equations along contour
tol = tolselect( cim, data, 'tol', 1e-2 );  %  select modes according to singular values
\end{code}
Upon execution of the \verb!tolselect! command, the window shown in the left panel of Fig.~\ref{fig:cimgui} opens up.  It displays the singular values returned from the \cim algorithm.  For typical problems, the resonance modes have the largest singular values and are separated by a sufficiently large gap from those modes which should be discarded because they do not noticeably contribute to the resonance mode expansion.  We will comment on this point in more detail in Sec.~\ref{sec:cimresults}.  The user can select the relevant modes interactively with the mouse cursor (red line) and by clicking the left mouse button.  Once the modes (red dots in left panel) are selected, we can visualize the resonance frequencies in the complex plane by pressing the ellipse icon in the \textsc{matlab} figure panel.  This opens the figure shown in the right panel of Fig.~\ref{fig:cimgui}.  To continue from here on with the \cim simulation, the user should press the green triangle icon $\triangleright$.  If something has gone wrong with the computation of the resonance modes, the user should press the red circle $\bullet$.  In the calling program the progress of the \cim calculation is then handled with
\begin{code}
if isempty( tol ),  return;  end
cim = eval2( cim, data, 'tol', tol );
\end{code}
The first line makes sure that the user has not pressed the red circle icon $\bullet$, otherwise the calculation is stopped.  In the second line we keep the modes with a sufficiently large singular value and normalize the modes, using the approach discussed in Sec.~\ref{sec:theory}.  Once the \cim solver has been initialized, it can be used in the same manner as the previously described \bem solver.  For instance, using the excitation object \verb!exc! for plane waves we can compute the \bem solution through
\begin{code}
sol = bem \ exc( tau, k0 );
\end{code}
This step is usually fast because all relevant information about the electromagnetic response to an external excitation is already contained in the resonance modes.  Yet, the description is often extremely accurate, as can be seen for instance by comparing the results of the \bem and \cim simulations in Figs.~\ref{fig:cim01} and \ref{fig:cim02}, which are almost indistinguishable.

\section{Theory}\label{sec:theory}

In this section, we first introduce to the theory of the boundary integral method approach and its numerical implementation using boundary elements, and then discuss the basic ideas behind resonance modes and resonance mode expansions.

\subsection{Boundary integral method}

Within the boundary integral method one considers one or several particles embedded in a background medium, as depicted in Fig.~\ref{fig:bemschem1}(a).  In the following we discuss for simplicity the case of a single particle only, but our approach can be easily generalized and the \nanobem toolbox implements general geometries with possibly coupled or coated particles.  Our main assumption about the system under study is that the electromagnetic response of the particle and the background medium is linear and can be described in terms of homogeneous, isotropic, and local material parameters.  We denote the permittivity and permeability of the particle with $\varepsilon_1$ and $\mu_1$, respectively, and the material properties of the background medium with $\varepsilon_2$, $\mu_2$.  The particle volume is $\Omega$, the particle boundary is $\partial\Omega$, and the outer surface normal pointing away from the particle is $\hat{\bm n}$.  The boundary integral method, to be discussed below, is a solution scheme for Maxwell's equations that only involves the tangential electromagnetic fields at the particle boundary $\bm s\in\partial\Omega$,
\begin{equation}
  \bm u_E(\bm s)=\hat{\bm n}\times\bm E(\bm s)\,,\qquad
  \bm u_H(\bm s)=\hat{\bm n}\times\bm H(\bm s)\,.
\end{equation}
We consider the case of \textit{incoming} electromagnetic fields $\bm E_2^{\rm inc}$, $\bm H_2^{\rm inc}$ oscillating with a given angular frequency $\omega$, which are either solutions of the homogeneous Maxwell's equations, e.g., plane waves, or are produced by current sources located within the background medium, e.g., oscillating dipoles.  These incoming fields would be the solutions of Maxwell's equations in absence of the particle.  In presence of the particle we additionally have to consider the scattered fields $\bm E_1^{\rm sca}$, $\bm H_1^{\rm sca}$ and $\bm E_2^{\rm sca}$, $\bm H_2^{\rm sca}$ inside and outside the particle, respectively, which have to be chosen such that the boundary conditions of Maxwell's equations hold at the particle boundary.  The total electromagnetic fields are the sum of incoming and scattered fields.  As discussed in Refs.~\cite{chew:95,hohenester:20} and briefly outlined in \ref{sec:representation}, for positions $\bm r$ inside the particle the electromagnetic fields can be obtained from the representation formulas (we use SI units throughout)
\begin{subequations}\label{eq:representation}
\begin{eqnarray}
  \bm E(\bm r) &=& -i\omega\mu_1\bigl[\mathbb{S}_1\,\bm u_H\bigr](\bm r)+
  \bigl[\mathbb{D}_1\,\bm u_E\bigr](\bm r)\nonumber\\
  \bm H(\bm r) &=& +i\omega\varepsilon_1\bigl[\mathbb{S}_1\,\bm u_E\bigr](\bm r)+
  \bigl[\mathbb{D}_1\,\bm u_H\bigr](\bm r)\,.\qquad
\end{eqnarray}
Similarly, for positions $\bm r$ outside the particle the electromagnetic fields can be obtained from
\begin{eqnarray}
  \bm E(\bm r) &=& \bm E_2^{\rm inc}(\bm r)+i\omega\mu_2\bigl[\mathbb{S}_2\,\bm u_H\bigr](\bm r)-
  \bigl[\mathbb{D}_2\,\bm u_E\bigr](\bm r)\nonumber\\
  \bm H(\bm r) &=& \bm H_2^{\rm inc}(\bm r)-i\omega\varepsilon_2\bigl[\mathbb{S}_2\,\bm u_E\bigr](\bm r)-
  \bigl[\mathbb{D}_2\,\bm u_H\bigr](\bm r)\,.\qquad
\end{eqnarray}
\end{subequations}
Here $\bigl[\mathbb{S}\,\bm u_{E,H}\bigr](\bm r)$ and $\bigl[\mathbb{D}\,\bm u_{E,H}\bigr](\bm r)$ are the single and double layer potentials, for details see~\ref{sec:representation}, which propagate the tangential electromagnetic fields $\bm u_{E,H}$ from the particle boundary to position $\bm r$.  Note that owing to the boundary conditions of Maxwell's equations $\bm u_{E,H}$ are the same inside and outside the particle.  Eq.~\eqref{eq:representation} is reminiscent of Huygen's principle which propagates the fields at the wavefront (here $\bm u_{E,H}$) to another position in space (here $\bm r$).  Thus, once $\bm u_{E,H}$ are known at the boundary, we can compute the electromagnetic fields everywhere else using the representation formulas of Eq.~\eqref{eq:representation}.

The tangential fields $\bm u_{E,H}$ can be obtained by letting $\bm r$ in Eq.~\eqref{eq:representation} approach the boundary of the particle, and by exploiting the boundary conditions of Maxwell's equations.  While this limit can be performed safely for the single layer potential, $\lim_{\bm r\to\bm s}\hat{\bm n}\times\bigl[\mathbb{S}\,\bm u_{E,H}\bigr](\bm r)=\hat{\bm n}\times\bigl[\mathbb{S}\,\bm u_{E,H}\bigr](\bm s)$, in the evaluation of the double layer potential we have to be careful on whether we approach the boundary from the inside or outside~\cite{hohenester:20}
\begin{equation}
  \lim_{\bm r\to\bm s}\hat{\bm n}\times\bigl[\mathbb{D}\,\bm u_{E,H}\bigr](\bm r)=
  \pm\mbox{$\frac 12$}\bm u_{E,H}(\bm s)+\hat{\bm n}\times\bigl[\mathbb{D}\,\bm u_{E,H}\bigr](\bm s)\,.
\end{equation}
Here the positive sign has to be taken for the limit from the inside, and the negative sign for the limit from the outside.  Thus, if we consider in Eq.~(\ref{eq:representation}a) the tangential fields $\hat{\bm n}\times\bm E$, $\hat{\bm n}\times\bm H$ and approach the boundary from the inside, we get
\begin{subequations}\label{eq:representation2}
\begin{eqnarray}
  \mbox{$\frac 12$}\bm u_E(\bm s) &=& -i\omega\mu_1\hat{\bm n}\times\bigl[\mathbb{S}_1\,\bm u_H\bigr](\bm s)+
  \hat{\bm n}\times\bigl[\mathbb{D}_1\,\bm u_E\bigr](\bm s)\nonumber\\
  \mbox{$\frac 12$}\bm u_H(\bm s) &=& +i\omega\varepsilon_1\hat{\bm n}\times\bigl[\mathbb{S}_1\,\bm u_E\bigr](\bm s)+
  \hat{\bm n}\times\bigl[\mathbb{D}_1\,\bm u_H\bigr](\bm s)\,.\qquad
\end{eqnarray}
Similarly, we obtain from Eq.~(\ref{eq:representation}b)
\begin{eqnarray}
  \mbox{$\frac 12$}\bm u_E(\bm s) &=& \hat{\bm n}\times\bm E_2^{\rm inc}(\bm s)+
  i\omega\mu_2\hat{\bm n}\times\bigl[\mathbb{S}_2\,\bm u_H\bigr](\bm s)-
  \hat{\bm n}\times\bigl[\mathbb{D}_2\,\bm u_E\bigr](\bm s)\nonumber\\
  \mbox{$\frac 12$}\bm u_H(\bm s) &=& \hat{\bm n}\times\bm H_2^{\rm inc}(\bm s)-
  i\omega\varepsilon_2\hat{\bm n}\times\bigl[\mathbb{S}_2\,\bm u_E\bigr](\bm s)-
  \hat{\bm n}\times\bigl[\mathbb{D}_2\,\bm u_H\bigr](\bm s)\,.\qquad
\end{eqnarray}
\end{subequations}
For given incoming fields $\bm E_2^{\rm inc}$, $\bm H_2^{\rm inc}$, we must extract from the two sets of equations~\eqref{eq:representation2} the unknown $\bm u_{E,H}$.  In the Poggio-Miller-Chang-Harrington-Wu-Tsai formulation~\cite{chang:77,poggio:73,wu:77} one subtracts the two sets of equations to arrive at~\cite{hohenester:20}
\begin{eqnarray}\label{eq:bim}
  \hat{\bm n}\times\Bigl({\mathbb{D}}_1[\bm u_E](\bm s)+{\mathbb{D}}_2[\bm u_E](\bm s)\Bigr)-
  \hat{\bm n}\times\Bigl(i\omega\mu_1{\mathbb{S}}_1[\bm u_H](\bm s)+i\omega\mu_2{\mathbb{S}}_2[\bm u_H](\bm s)\Bigr) &=&
  \hat{\bm n}\times\bm E_2^{\rm inc}(\bm s) \nonumber\\
  \hat{\bm n}\times\Bigl({\mathbb{D}}_1[\bm u_H](\bm s)+{\mathbb{D}}_2[\bm u_H](\bm s)\Bigr)+
  \hat{\bm n}\times\Bigl(i\omega\varepsilon_1{\mathbb{S}}_1[\bm u_E](\bm s)+i\omega\varepsilon_2{\mathbb{S}}_2[\bm u_E](\bm s)\Bigr) &=&
  \hat{\bm n}\times\bm H_2^{\rm inc}(\bm s)\,.
\end{eqnarray}
These are the working equations for the boundary integral method approach.

\subsection{Boundary element method}\label{sec:bem}

\begin{figure}
\centerline{\includegraphics[width=0.95\textwidth]{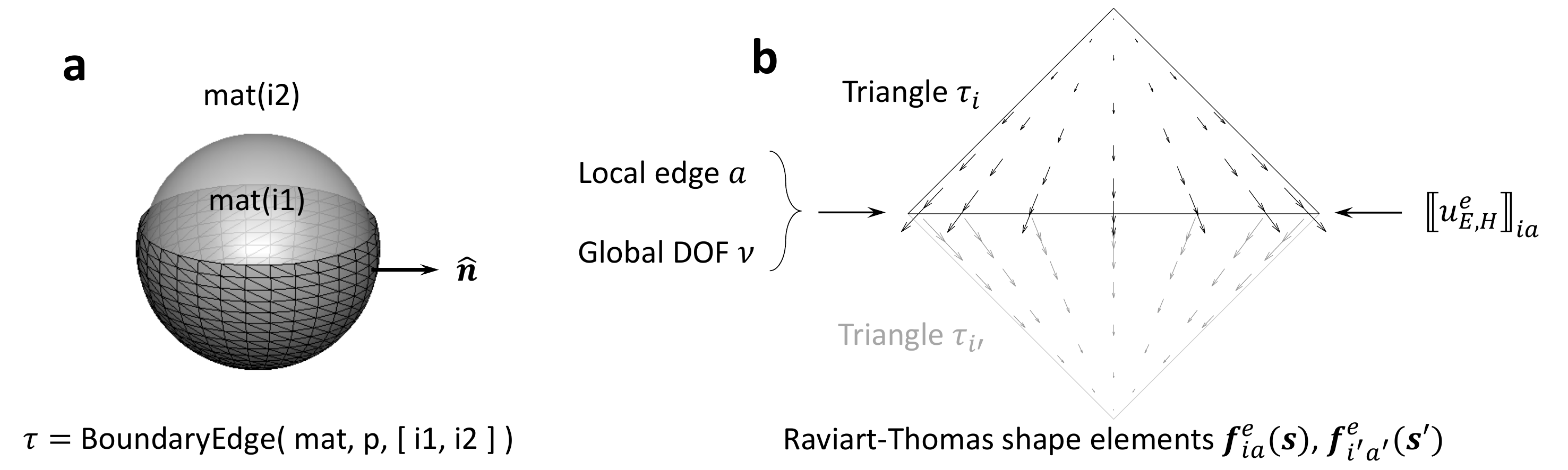}}
\caption{Discretized particle boundary and Raviart-Thomas shape elements for boundary element method (\bem) approach.  (a) In \bem, the nanoparticle boundary is discretized by boundary elements, here of triangular shape.  (b) In each triangle $\tau_i$ we introduce Raviart-Thomas shape elements $f_{ia}(\bm s)$ for the representation of the tangential electromagnetic fields.  The shape elements are constructed such that the inflow to an edge equals the outflow to the adjacent triangle.  For the representation of the tangential fields we attach a value $\mmatrix[\normalsize]{u_{E,H}^e}$ to each edge and interpolate the fields in the triangles using the shape elements.  Each edge is labeled by a unique global degree of freedom $\nu$, within each triangle we use the local degrees of freedom $i,a$, where $a$ denotes the different edges of triangle $\tau_i$.}\label{fig:bemschem02}
\end{figure}

The boundary element method (\bem) approach implements Eq.~\eqref{eq:bim} for a discretized particle boundary
\begin{equation}
  \partial\Omega=\bigcup_{i=1}^m\tau_i\,,
\end{equation}
where $\tau_i$ denote the different boundary elements and $m$ is the total number of elements.  See also Fig.~\ref{fig:bemschem02}.  In the \nanobem toolbox we currently provide boundary elements of triangular shape only and consider regular mesh triangulations where each edge of a given triangle is shared by a single adjacent triangle.  In addition to the discretized particle boundary, we also need a discretization for the tangential electromagnetic field $\bm u_{E,H}$.  A convenient choice are the Raviart-Thomas or Rao-Wilton-Glison basis elements, which guarantee the continuity of the tangential fields when going from one triangle to an adjacent triangle~\cite{chew:95,hohenester:20}.  Technically, this is done by assigning to each edge $\nu$ of the discretized particle a value for $\mmatrix[\normalsize]{u_{E,H}^e}_\nu$ and by using basis functions $\bm f_\nu^e(\bm s)$ that are nonzero in the two adjacent triangles only (a so-called local support), and which are constructed such that the outflow from one triangle equals the inflow to the other triangle.  For details see~\cite{chew:95,hohenester:20}.  The discretized tangential fields can then be approximated through
\begin{equation}\label{eq:shape}
  \bm u_{E,H}(\bm s)=\sum_{\nu=1}^n \mmatrix[\big]{u_{E,H}^e}_\nu\,\bm f^e_\nu(\bm s)\,,
\end{equation}
where $n$ is the total number of individual edges that equals the number of degrees of freedom for the \bem approach. Finally, Eq.~\eqref{eq:bim} is converted into a matrix equation using a Galerkin scheme.  For the single layer potential we obtain~\cite{hohenester:20}
\begin{equation}\label{eq:singlebem}
  \mmatrix[\big]{\mathbb S}_{\nu\nu'}=\oint \bm f^e_{\nu}(\bm s)\cdot\big[\mathbb S\,\bm f^e_{\nu'}\big](\bm s)\,
  dS\,,
\end{equation}
where $dS$ denotes the integration over the boundary, with a similar expression for the double layer potential.  Note that because of the local support of the basis function $\bm f_\nu^e$ the above integral can be broken down into integrals over triangle pairs.  Finally, the working equations~\eqref{eq:bim} of the boundary integral method approach can be approximated through a matrix equation
\begin{equation}\label{eq:bem}
  \begin{pmatrix} 
    \mmatrix[\normalsize]{{\mathbb{D}}_1+{\mathbb{D}}_2} &
    -i\omega\mmatrix[\normalsize]{\mu_1{\mathbb{S}}_1+\mu_2{\mathbb{S}}_2} \\
    i\omega\mmatrix[\normalsize]{\varepsilon_1{\mathbb{S}}_1+\varepsilon_2{\mathbb{S}}_2} &
    \mmatrix[\normalsize]{{\mathbb{D}}_1+{\mathbb{D}}_2}
  \end{pmatrix}
  \left(\begin{array}{c} \mmatrix[\normalsize]{u_E^e} \\ \mmatrix[\normalsize]{u_H^e} \end{array}\right)=
  \left(\begin{array}{c} \mmatrix[\normalsize]{q_E^{\rm inc}} \\ \mmatrix[\normalsize]{q_H^{\rm inc}} \end{array}\right)\,,
\end{equation}
with the inhomogeneity of the incoming fields
\begin{equation}\label{eq:qinc}
  \mmatrix{q_{E,H}^{\rm inc}}_\nu=\oint\bm f_{\nu}^e(\bm s)\cdot
  \left\{\begin{array}{c}\bm E^{\rm inc}(\bm s) \\ \bm H^{\rm inc}(\bm s) \end{array}\right\}\,dS\,.
\end{equation}
Details of the implementation of Eq.~\eqref{eq:bem} within the \nanobem toolbox will be given below.

\subsection{Resonance modes and contour integral method}

In many cases of interest, the working equation~\eqref{eq:bem} can be approximately solved by using resonance modes.  We start by rewriting the working equations in the more compact form
\begin{equation}\label{eq:bem2}
  \begin{pmatrix} A_{11} & A_{12} \\ A_{21} & A_{22} \\ \end{pmatrix}
  \begin{pmatrix} u_E \\ u_H \\ \end{pmatrix} = \begin{pmatrix} q_E \\ q_H \\ \end{pmatrix}\,,
\end{equation}
or $A(\omega) u=q$ in short, where we have explictly indicated the dependence of the transmission matrix $A$ on the angular frequency $\omega$ of the external excitation.  A resonance mode is characterized through a (in general complex) resonance frequency $\omega_k$ for which the following expression holds
\begin{equation}\label{eq:resonancemode}
  A(\omega_k) v_k=0\,,\qquad v_k=\begin{pmatrix} v_E \\ v_H \\ \end{pmatrix}\,.
\end{equation}
Here $v_k$ is the resonance mode associated with $\omega_k$.  Due to our formulation of the \bem equations, see also \ref{sec:representation}, it is guaranteed that these modes have the proper boundary conditions of outgoing waves at infinity.  The computation of the resonance modes is particularly simple within the \bem approach, contrary to many other computational schemes such as the finite element method approach, where the consideration of complex frequencies requires a careful treatment of the perfectly matched layers in order to properly damp the outgoing waves.  In contrast, within \bem the single and double layer potentials only connect positions on the boundary of the particle, and $\mathbb{S}$, $\mathbb{D}$ can be safely evaluated for complex frequencies.  In the \nanobem toolbox we employ the contour integral method approach of Beyn~\cite{beyn:12}, which allows computing the eigenmodes in a complete black box manner: we solve Eq.~\eqref{eq:bem2} for random inhomogeneities along a contour in the complex frequency plane, and the algorithm provides the resonances located inside the contour.  Details of our implementation of the contour integral method (\cim) and its working principles will be presented in Sec.~\ref{sec:cimsolver}.

In addition to the right eigenmodes of Eq.~\eqref{eq:resonancemode} there also exist left eigenmodes $\tilde\nu_k$ defined through  
\begin{equation}\label{eq:leftresonancemode}
  \tilde v_k^T A(\omega_k)=0\,,\qquad \tilde v_k=\begin{pmatrix} v_H \\ v_E \\ \end{pmatrix}\,,
\end{equation}
where $\tilde v_k$ is obtained from $v_k$ by exchanging the electric and magnetic components.  This can be easily proven by taking the transpose of Eq.~\eqref{eq:resonancemode} and noting that all of the submatrices $A_{11}$, $A_{12}$, $A_{21}$, $A_{22}$ in Eq.~\eqref{eq:bem2} are symmetric, and additonally $A_{11}=A_{22}$ holds, as can be inferred from the matrix in Eq.~\eqref{eq:bem}. 

In a second step, we approach the reconstruction problem and seek for a description of the electromagnetic response to some incoming fields in terms of resonance modes via
\begin{equation}\label{eq:cimansatz}
  u_{\rm CIM}(\omega)\approx\sum_{k=1}^K C_k(\omega)\,v_k\,,
\end{equation} 
where $K$ is the number of modes under consideration, e.g., those located inside a given contour in the complex frequency plane, and $C_k(\omega)$ are the expansion coefficients to be determined.  In case of a single mode $v_k$, we can expand close to the resonance $\omega\approx\omega_k$ the transmission matrix into a Taylor series and rewrite the \bem equation $A(\omega)u=q$ approximately as
\begin{equation}
  \Big(A(\omega_k)+A'(\omega_k)\big(\omega-\omega_k)\Big)C_k(\omega)v_k\approx q\,,
\end{equation}
where $A'(\omega_k)$ denotes the derivative of the transmission matrix at the resonance frequency.  Multiplying the equation from the left-hand side with $\tilde v_k^T$ and using Eq.~\eqref{eq:leftresonancemode}, allows us to express the expansion coefficient as
\begin{equation}
  C_k(\omega)\approx \Big(\tilde v_k^T A'(\omega_k) v_k\Big)^{-1}
  \frac{\tilde v_k^T q}{\omega-\omega_k}\,.
\end{equation}
The term in parentheses is a normalization constant, and the fraction is a Lorentzian lineshape function with the resonance frequency $\omega_k$ and the oscillator strength $\tilde v_k^Tq$.  In case of many resonance modes, one can show that the expansion of Eq.~\eqref{eq:cimansatz} can be written in the form
\begin{equation}\label{eq:cim}
  u_{\rm CIM}(\omega)=\sum_{k=1}^K \Big(\tilde v_k^T A'(\omega_k) v_k\Big)^{-1}
  \frac{\tilde v_k^T q}{\omega-\omega_k}v_k+h(\omega)\,,
\end{equation}
where $h(\omega)$ is a holomorphic function that contains contributions from resonance modes located outside the contour and terms originating from higher-order contributions of the Taylor expansion of the transmission matrix $A(\omega)$.  Eq.~\eqref{eq:cim} follows from Keldysh's theorem~\cite{kozlov:99,mennicken:03}, a central result of the theory of eigenvalue problems for holomorphic Fredholm operator-valued functions, which can be directly applied to our \bem formulation.  In many cases of interest, the contribution of $h(\omega)$ can be neglected, but there is no guarantee that such an approach always works.  See Sec.~\ref{sec:cimsolver} for a discussion of this issue and the possibility to consider a non-resonant background.

\section{The \nanobem toolbox}\label{sec:toolbox}

The \nanobem toolbox is a \textsc{matlab} implementation of the Galerkin scheme for the boundary element method approach using Raviart-Thomas basis functions.  It is written in an object oriented manner with the main emphasis on clarity rather than speed.  We have currently tested the code for relatively small problems only, say a few thousand boundary elements.  In the following we discuss the various steps of a typical \bem simulation in more detail, which require setting up the particle boundaries and the dielectric environment, specifying the external perturbation, and solving the \bem equations, with a possible post processing of the final solution vector.  Note that the toolbox comes along with \textsc{matlab} help pages that provide further information about the implementation details.  Table~\ref{tab:nanobem} lists the most important classes of the toolbox.

\begin{table}
\caption{List of the most important classes of the \nanobem toolbox.  A detailed description of the various classes is given in Sec.~\ref{sec:toolbox} and in \ref{sec:details}.}\label{tab:nanobem}
{\footnotesize
\begin{tabularx}{\columnwidth}{lX}
\hline\hline
Class name & Description \\
\hline
\texttt{Material} & Material properties of media \\
\texttt{particle} & Discretization of particle boundary in terms of vertices and faces \\
\texttt{BoundaryEdge} & Boundary elements with Raviart-Thomas shape elements \\
\texttt{Point} & Point embedded in photonic environment \\[4pt]
\texttt{galerkin.planewave} & Planewave excitation \\
\texttt{galerkin.dipole} & Oscillating dipole excitation \\
\texttt{galerkin.bemsolver} & \bem solver for solution of working equation \eqref{eq:bem} \\
\texttt{galerkin.cimsolver} & Resonance mode solver based on the contour integral method (\cim) \\
\texttt{galerkin.solution} & Solution vector for $\mmatrix[\normalsize]{u_{E,H}}$ of Eq.~\eqref{eq:shape} \\
\hline
\hline
\end{tabularx}}
\end{table}

\subsection{Units}

With exception of the \texttt{Material} class and the \cim solver, in the \nanobem toolbox lengths are measured in nanometers and frequencies are converted to the corresponding wavenumbers of light in vacuum
\begin{equation}
  k_0=\frac{2\pi}\lambda=\frac\omega c\,.
\end{equation}
$\lambda$ is the wavelength of light in vacuum, $\omega$ the angular frequency, and $c$ the speed of light.  Occasionally one would like to convert from eV to wavenumbers.  This can be done through
\begin{code}
ene = 2;                  %  photon energy in eV
units;                    %  load conversion factor from eV to wavelength in nm
lambda = eV2nm / ene;     %  convert to wavelength in nm
k0 = 2 * pi / lambda;     %  convert wavelength to wavenumber
\end{code}
Electric fields are given in units of the field strength of the incoming field, which can be also set to one, such that the simulations give the field enhancement only.  Instead of the magnetic field we internally use $Z_0H$, with $Z_0$ being the free space impedance, such that the electric field and the scaled magnetic field  have the same units.

\subsection{Material class}

The material properties of a medium are implemented in the \nanobem toolbox through the \texttt{Material} class
\begin{code}
mat = Material( eps, mu );    %  initialize Material object
\end{code}
Here \texttt{eps} is either a constant or a permittivity function \verb!eps(ene)! with the photon energy in eV as input, with a corresponding functionality for the permeability \verb!mu!.  Both \verb!eps! and \verb!mu! should return dimensionless numbers.  We provide functions for tabulated and Drude dielectric functions
\begin{code}
eps = epstable( 'gold.dat' );   %  use tabulated dielectric function
eps = epsdrude( 'Au' );         %  Drude dielectric function
\end{code}
Once the material object is initialized, we can compute for a given free space wavenumber~\verb!k0! the permittivity \verb!mat.eps(k0)!, the permeability \verb!mat.mu(k0)!, the effective wavenumber \verb!mat.k(k0)!, and the impedance \verb!mat.Z(k0)!.  In all \nanobem simulations we first have to set up a table of material properties, e.g., for air and silver
\begin{code}
mat1 = Material( 1, 1 );                         %  air
mat2 = Material( epstable( 'silver.dat' ), 1 );  %  silver
mat = [ mat1, mat2 ];                            %  unique material vector
\end{code}
The material vector \texttt{mat} should be specified once, and then the same vector should be passed to all objects and functions of the toolbox.  We recommend setting the first entry of the material vector to the material of the background medium.

\subsection{Boundary}

In a typical \bem simulation we deal with several connected or unconnected boundaries
\begin{equation}
  \partial\Omega=\bigcup_{j=1}^n\partial\Omega_j\,,
\end{equation}
where each boundary $\partial\Omega_j$ has an orientation, this is an inside and outside specified with respect   to the outer surface normal $\hat{\bm n}$, and each boundary $\partial\Omega_j$ must have a single material at the inside and outside.  The boundaries $\partial\Omega_j$ are discretized using triangles $\tau_i$ through
\begin{equation}
  \partial\Omega_j=\bigcup_{i=1}^{m}\tau_i\,,
\end{equation}
where $m$ is the total number of triangles.  In the \nanobem toolbox these boundaries are specified using the \verb!particle! class
\begin{code}
p = particle( verts, faces );   %  set up particle boundary
\end{code}
Here \verb!verts! are the triangle corners or vertices, and \verb!faces! is a $m\times 3$ matrix describing how these vertices are connected within the individual triangles.  Note that this implementation is in complete analogy to the face-vertex structures provided by \textsc{matlab}.  The toolbox provides a few implementations for simple shapes, such as spheres, rods, or cubes, together with a plotting function for the particle boundary.  Details can be found in the help pages of the toolbox.  For more complicated setups the user must provide a discretization of the boundaries of the desired structure in terms of vertices and triangular faces, which often constitutes the most difficult part in setting up a \bem simulation.

\subsection{Shape elements}

Once we have set up the material vector and the boundaries, we can initialize the Raviart-Thomas shape elements which play a central role in our \bem approach,
\begin{code}
tau = BoundaryEdge( mat, p, inout );   
tau = BoundaryEdge( mat, p1, inout1, p2, inout2, ... );
\end{code}
Here \verb!mat! is a unique material vector, \verb!p! is a \texttt{particle} object characterizing the discretized particle boundary, and \verb!inout! is a vector \verb![i1,i2]! where \verb!mat(i1)! is the material at the particle inside and \verb!mat(i2)! the material at the particle outside.  See also Fig.~\ref{fig:bemschem02}.  For example, to set up a silver sphere embedded in glass (refractive index 1.5) we proceed as follows
\begin{code}
mat = [ Material( 1.5 ^ 2, 1 ), Material( epstable( 'silver.dat' ), 1 ) ];
diameter = 50;
p = trisphere( 144, diameter );
tau = BoundaryEdge( mat, p, [ 2, 1 ] );
\end{code}
In the first line we set up the material vector, in the third line we define a sphere triangulation with 144 vertices (diameter 50 nm), using the toolbox function \verb!trisphere!, and in the last line we set up the shape elements for the nanosphere boundary with silver at the inside and glass at the outside.  \verb!tau! is a \texttt{BoundaryEdge} vector specifying the shape elements for the discretized particle boundary, where each vector element has the following properties
\begin{code}
tau(i).nu;         %  global degrees of freedom for triangle edges
tau(i).val;        %  auxiliary information for Thomas-Raviart basis elements
tau(i).inout;      %  index to material at inside and outside
tau(i).verts;      %  triangle vertices
\end{code}
\verb!nu! are the three global degrees of freedom for the shape elements assigned to the three edges of the triangle, see Eq.~\eqref{eq:shape} and Fig.~\ref{fig:bemschem02}, and \verb!val! is a value for the Raviart-Thomas elements that is chosen such that the flow of $u_{E,H}$ in and out of the triangle edges is conserved.  \verb!inout! specifies the material at the inside and outside of the triangle boundary, and \verb!verts! are the triangle corners.  Note that in our present implementation we internally assign the shape elements to the boundary elements, and evaluate the single and double layer potentials for triangle pairs rather than shape element pairs, which extend over two triangles.  This allows for some speedup for discretizations with a moderate number of boundary elements, say a few thousands, but may become inefficient for larger discretizations.  From here on all information about the discretized boundary is stored in the \verb!BoundaryEdge! vector, the information of the \verb!particle! object is no longer needed.  We note in passing that the toolbox also implements linear shape elements \verb!BoundaryVert! for simulations using the quasistatic approach, for details see the help pages of the toolbox.

\subsection{Plane wave and dipole excitation}

At present, the \nanobem toolbox provides two types of external excitations, namely plane waves and oscillating dipoles.  Other excitations can be easily implemented, as described in more detail in \ref{sec:details}, all a user must provide is a structure of the form
\begin{code}
  qinc = struct( 'e', e, 'h', h, 'tau', tau, 'k0', k0 );
\end{code}
Here \verb!e!, \verb!h! are the inhomogeneities given in Eq.~\eqref{eq:qinc}, \verb!tau! is a vector for the boundary elements, and \verb!k0! is the free space wavenumber of the incoming fields.  The dimensions of \verb!e!, \verb!h! must be $n\times 1$ for a single excitation, where $n$ is the total number of edges or degrees of freedom, see Eq.~\eqref{eq:shape}.  It is also possible to specify multiple excitations, e.g. different polarization vectors, by providing array dimensions of $n\times m$ or any other multidimensional shape with leading dimension $n$.  For plane wave excitations, we first set up a \verb!galerkin.planewave! object which is then evaluated for a given bounday \verb!tau! and wavenumber \verb!k0!
\begin{code}
exc = galerkin.planewave( [ 1, 0, 0; 0, 1, 0 ], [ 0, 0, 1; 0, 0, 1 ] );
qinc = exc( tau, k0 );
\end{code}
The first array in the call to \verb!galerkin.planewave! gives the polarization vectors, here polarization along $x$ and $y$, and the second array gives the light propagation direction, here along~$z$.  To set up an excitation of oscillating dipoles, we must first specify the positions of the dipoles and in which medium they are located
\begin{code}
pt = Point( mat, imat, pos );
\end{code}
\verb!mat! is the same material vector that is passed to \verb!BoundaryEdge!, \verb!imat! specifies the medium (scalar) or media (vector) in which the dipole positions \verb!pos! are located.  Alternatively, the positions can be also placed automatically in the photonic environment specified by \verb!tau!
\begin{code}
pt = Point( tau, pos );
\end{code}
Once the \verb!Point! object is initialized, we can set up and evaluate the \verb!galerkin.dipole! object in the same way as for plane waves
\begin{code}
dip = galerkin.dipole( pt );
qinc = dip( tau, k0 );
\end{code}
\verb!e!, \verb!h! are arrays of dimension $n\times m_{\rm pt}\times 3$, where $m_{\rm pt}$ is the number of dipole positions and the last dimension accounts for the dipole orientations along $x$, $y$, and $z$.

\subsection{The boundary element method solver}

To solve the \bem equations~\eqref{eq:bem}, one has to set up a \verb!galerkin.bemsolver! object and invert the matrix equation for the excitation structure \verb!qinc!
\begin{code}
bem = galerkin.bemsolver( tau );
sol = bem \ qinc;
\end{code}
Both the initialization and the solution of the \bem solver are typically the most time consuming parts of a simulation.  The performance and the working principle of the \bem solver can be controlled through additional arguments passed in the initialization call to the solver, as described in more detail in \ref{sec:bemdetails} and the help pages of the toolbox.  The \bem solution \verb!sol! is an object of type
\begin{code}
sol = galerkin.solution( tau, k0, e, h );
\end{code}
where \verb!tau! is the discretized boundary, \verb!k0! is the free space wavenumber for the external excitation, and \verb!e!, \verb!h! hold the values $\mmatrix[\normalsize]{u_{E,H}^e}$ characterizing the tangential electromagnetic fields of Eq.~\eqref{eq:shape}.  The \bem solution object \verb!sol! can be used for various kinds of post processing.  For a planewave excitation we can pass the solution to the planewave object in order to compute the optical cross sections
\begin{code}
csca = scattering( exc, sol );  %  scattering cross section in (nm^2)
cabs = absorption( exc, sol );  %  absorption cross section in (nm^2)
cext = extinction( exc, sol );  %  extinction cross section in (nm^2)
\end{code}
For a dipole excitation, we can use the dipole object in order to compute the total and radiative decay rates in units of the free-space decay rate~\cite[chapter 10]{hohenester:20}
\begin{code}
[ tot, rad ] = decayrate( dip, sol );
\end{code}
For all types of excitations we can compute the electromagnetic fields at the particle boundary and the surface charge distribution 
\begin{code}
[ e1, h1 ] = interp( sol, 'inout', 1 );  %  fields at boundary inside
[ e2, h2 ] = interp( sol, 'inout', 2 );  %  fields at boundary outside
sig = surfc( sol );  %  surface charge distribution
\end{code}
The fields and the surface charge distribution are given at the centroids of the boundary elements.  The toolbox provides a number of plotting routines for the fields and the surface charge distribution, as described in more detail in the help pages.  In order to evaluate the fields away from the boundary, we can use the representation formulas~\eqref{eq:representation}.  To this end we set up a \verb!Point! object and pass it together with the solution vector to the \verb!fields! function
\begin{code}
[ x, y ] = ndgrid( len * linspace( -0.5, 0.5, n ) );
pt = Point( tau, [ x( : ), y( : ), 0 * x( : ) ] );
[ e, h ] = fields( sol, pt );        %  evaluate scattered fields
[ ei, hi ] = fields( exc, pt, k0 );  %  evaluate incoming fields
\end{code}
In the first line we set up a regular grid of size \verb!len! with $\mbox{\tt n}\times\mbox{\tt n}$ points centered around the origin, and place in the second line the points into the photonic environment of \verb!tau!.  In the third and fourth line we compute the scattered and incoming fields, which have the dimension of $\mbox{\tt npt}\times 3\times\mbox{\tt m}$, where \verb!npt! is the number of points and \verb!m! the number of excitations.  Finally, we can also compute the optical far fields through
\begin{code}
[ efar, hfar ] = farfields( sol, pt ); 
\end{code}
The points \verb!pt! should be placed in the embedding medium and give the direction (measured with respect to the origin) along which the fields propagate far away from the nanoparticle.

\subsection{The contour integral method solver}\label{sec:cimsolver}

The contour integral method of Beyn~\cite{beyn:12} allows to determine the resonance frequencies $\omega_k$ and modes $v_k$ of Eq.~\eqref{eq:resonancemode} that are located inside a given closed contour $\Gamma$ in the complex frequency plane (see also Ref.~\cite{bykov:13} for a related scheme).  We start by briefly summarizing the method of Beyn~\cite{beyn:12} for the determination of the modes fulfilling $A(\omega_k)v_k=0$.  First, we set up a random matrix $V_r$ of dimension $n\times n_r$, where $n$ is the matrix size of $A(\omega)$ (this is the total number of global degrees of freedom for representing the electric and magnetic fields) and $n_r$ is a value that has to be larger than the number of resonance modes in the contour (which can only be determined at the end of the algorithm).  We usually set $n_r$ to a value of a few hundreds.  In a second step, we compute the matrices
\begin{equation}\label{eq:cimwork1}
  A_j=\frac 1{2\pi i}\oint_\Gamma z^jA^{-1}(z)V_r\,dz
\end{equation}
for different values of $j$.  We use an elliptic contour $\Gamma$ and approximate the contour integration through a simple trapezoidal quadrature.  The solution of $A(z)u=V_r$ for $u$ in Eq.~\eqref{eq:cimwork1} requires the solution of the \bem equation, which is typically the time consuming part of \cim (together with the computation of the normalization constants to be described at the end).  In the following we describe the procedure for obtaining the resonance modes from the knowledge of the matrices $A_0$, $A_1$, but we have also implemented the algorithm using higher-order matrices along the lines discussed in the paper of Beyn.  Quite generally, in all our computer experiments the \cim approach worked perfectly when using $A_0$, $A_1$ only.  In a third step, we perform a singular value decomposition of $A_0=V_0\Sigma_0 W_0^\dagger$ and keep only singular values that are larger than some cutoff value.  In the toolbox, this step is implemented through the graphical user interface described in Sec.~\ref{sec:start}, and we will discuss this step in more detail below in Sec.~\ref{sec:cimresults}.  In the fourth and last step, we compute the resonance frequencies $\omega_k$ and modes $v_k$ from the solutions of the eigenvalue problem
\begin{equation}\label{eq:cimwork2}
  \left(V_0^\dagger A_1W_0\Sigma_0^{-1}\right)s_k=\omega_k s_k\,,\quad v_k=V_0s_k\,.
\end{equation}
Once the resonance modes have been computed, we can check the accuracy of the numerical scheme by computing the residuals
\begin{equation}\label{eq:residua}
  r_k=\left\|A(\omega_k)v_k\right\|\,.
\end{equation}
Finally, we compute the left modes $\tilde v_k$ from Eq.~\eqref{eq:leftresonancemode} and the normalization factors in Eq.~\eqref{eq:cim}.  Inside the toolbox these normalization factors are absorbed into $\tilde v_k$.  The resonance modes and the contour in the complex frequency plane are given in units of the photon energy $\hbar\omega$ in eV.  The \cim solver is set up with
\begin{code}
cim = galerkin.cimsolver( bem );  %  use previously initialized BEM solver
cim = galerkin.cimsolver( tau );  %  or set up CIM solver using boundary elements
\end{code}
Once the \cim solver is initialized, we can set the size $n_r$ for the random matrix, the order \verb!nz! for \cim, and the contour $\Gamma$ through
\begin{code}
cim.nr = 200;   %  size of random matrix
cim.nz = 1;     %  order for CIM, default value is 1
cim.contour = cimbase.ellipse( [ zmin, zmax ], irad, nt );
\end{code}
In the initialization of the contour, \verb!zmin! and \verb!zmax! are the minimum and maximum of the contour on the real axis, \verb!irad! is the extension of the contour along the imagainary axis, and \verb!nt! is the number of quadrature points along the contour.  If needed, users can also shift the ellipse into the complex plane by adding an imaginary part to the first argument \verb![zmin,zmax]+1i*zshift!.  We next call
\begin{code}
data = eval1( cim, PropertyPairs );
\end{code}
In this call the \bem solver is initialized if we have previously set up the \cim solver with the boundary elements, where the performance of the \bem solver can be controlled with the same property pairs as in the initialization of the \bem solver, see also \ref{sec:details}, and additionally the matrices $A_j$ of Eq.~\eqref{eq:cimwork1} are evaluated along the discretized contour.  This evaluation can be rather time consuming depending on the number of boundary elements \verb!tau! and quadrature points.  The cutoff value for the singular values can be set interactively by the user with the \verb!tolselect! function described in Sec.~\ref{sec:start}, 
\begin{code}
tol = tolselect( cim, data );
if isempty( tol ),  return;  end
cim = eval2( cim, data, 'tol', tol );
\end{code}
The second line is needed for the case that the user stops the simulation in the call to \verb!tolselect!.  In the call to \verb!eval2! we compute the left resonance modes and normalization constants, which can be again time consuming because of the evaluation of $A'(\omega_k)$ for the normalization constant.  The \verb!cim! object now stores the following properties
\begin{code}
cim.nev     %  number of resonance modes
cim.ene     %  complex resonance frequencies
cim.sol     %  right resonance modes 
cim.sol2    %  left  resonance modes
cim.res     %  residuals
\end{code}
From here on the \cim solver can be used in the same manner as the \bem solver.  For instance, for given inhomogeneities \verb!qinc!, such as those of a planewave or dipole excitation, the approximate solution $u_{\rm CIM}$ of Eq.~\eqref{eq:cim} can be obtained through
\begin{code}
sol = cim \ qinc;
\end{code}
The outcome of this call is a \verb!galerkin.solution! vector that can be used for post processing purposes in exactly the same way as previously described in Sec.~\ref{sec:bem}.

For larger nanoparticles one occasionally observes a consistent deviation between the \bem and \cim results, see for instance the extinction spectra in Fig.~\ref{fig:cim02}(a,b) or Fig.~\ref{fig:bempp}(b,c).  As discussed for instance in \cite{zschiedrich:18,weiss:18,colom:18}, a spectrally broad, featureless and non-resonant background can account for this discrepancy.  We will not enter into this topic here, but present a simple approach for considering such a non-resonant background.  In the spirit of Eq.~\eqref{eq:cim}, we associate the difference between the inverse of the \bem transmission matrix $A^{-1}(\omega)$ and its \cim approximate as the holomorphic matrix
\begin{equation}\label{eq:holomorphic}
  H(\omega)=A^{-1}(\omega)-\sum_{k=1}^K \Big(\tilde v_k^T A'(\omega_k) v_k\Big)^{-1}
  \frac{v_k\tilde v_k^T}{\omega-\omega_k}\,,
\end{equation}
which accounts for the background.  As $H(\omega)$ is expected to have a weak frequency dependence, it usually suffices to compute $H(\omega)$ at a few sampling points only and to interpolate the function in between using a quadratic or cubic fit.  For instance, in Fig.~\ref{fig:cim02}(c) the three sampling energies are indicated by arrows in the bottom of the figure.  The spectral contribution of this background is shown by the dashed line, and by adding it to the \cim results one obtains perfect agreement with the full \bem results (compare circles and cross symbols).  To add a non-resonant background in the \nanobem toolbox, one first provides the sampling wavenumbers \verb!ktab! and then adds the contributions of Eq.~\eqref{eq:holomorphic} to the \cim solver via
\begin{code}
%  sampling wavenumbers for interpolation of nonresonant matrix
ktab = 2 * pi ./ [ 400, 600, 700 ];  
%  add nonresonant background to CIM solver, use BEM solver stored e.g. in data.bem
cim.nonresonant = diffcalderon( cim, bem, ktab );  
\end{code}
Once the non-resonant part has been added, the approximate \cim solution can again be obtained through {\small\verb!sol=cim\qinc!}.

\section{Frequently asked questions about resonance modes}\label{sec:cimresults}

\begin{figure}[t]
\includegraphics[width=\textwidth]{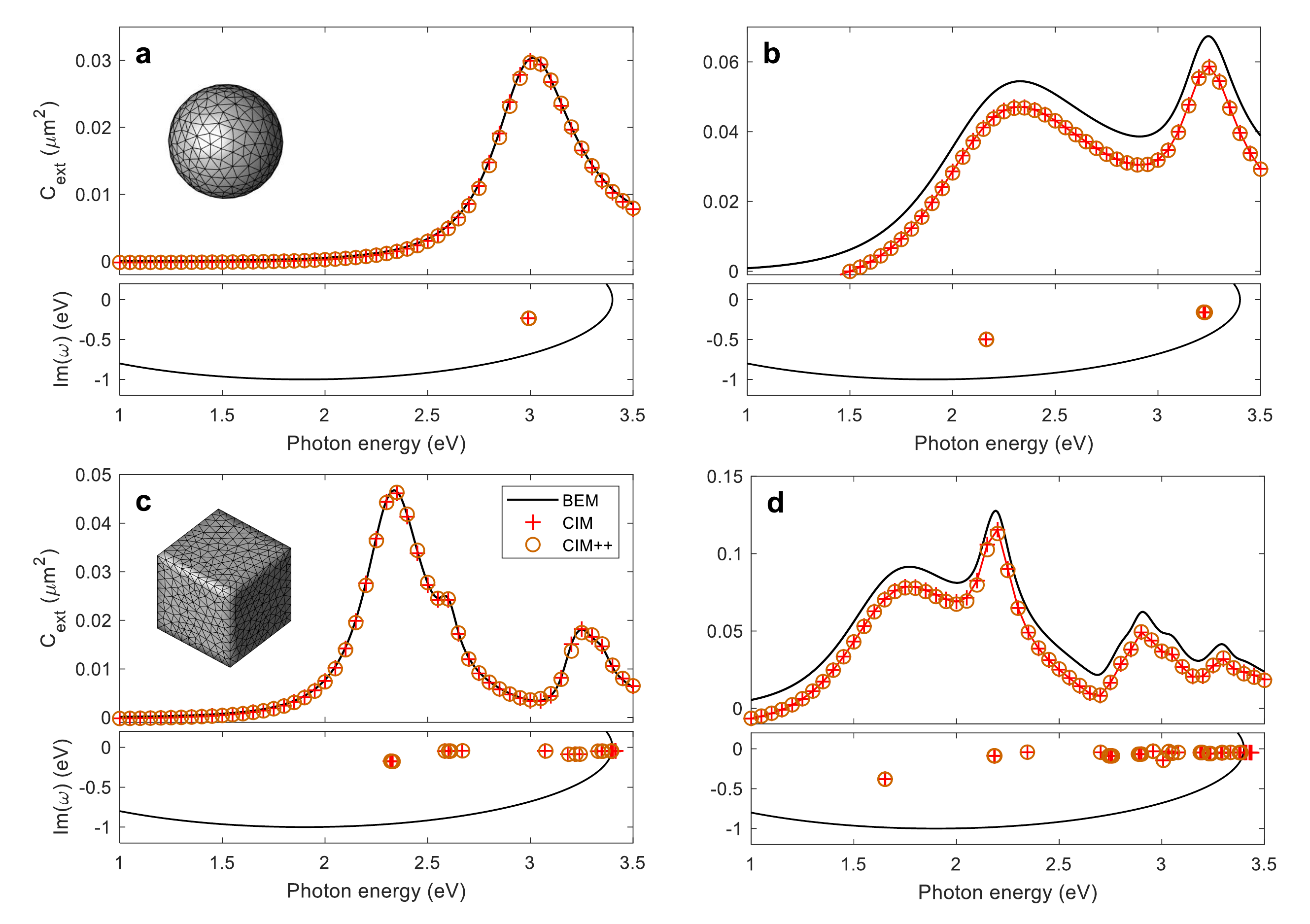}
\caption{Comparison of results obtained with the \nanobem toolbox and the \textsc{bem}++ software~\cite{bempp:15}.  We consider gold nanoparticles (Drude dielectric function taken from~\cite{perrin:16,unger:18}) embedded in glass (refractive index $n=1.5$), and use identical boundary discretizations and \cim contours in the \nanobem and \textsc{bem}++ simulations.  The black lines and the cross symbols ($+$) give the results of the \bem and \cim simulations obtained with the \nanobem toolbox, the circle symbols ($\circ$) give the results of the \cim simulations obtained with the \textsc{bem}++ software.  The different panels show results for a nanosphere with a diameter of (a) 50 nm and (b) 100 nm (305 vertices), and a nanocube with a side length of (c) 50 nm (153 vertices) and (d) 100 nm (611 vertices).  The upper parts of the panels show the extinction cross sections $C_{\rm ext}$, the lower parts display the contour for the determination of the resonance modes and the resonance energies $\hbar\omega_k$.  The \nanobem and \textsc{bem}++ results are in perfect agreement throughout.  The slight but systematic disagreement between the \bem and \cim results for the larger structures in panels (b,d) is due to the approximate nature of \cim, as discussed in more detail in~\cite{unger:18}, which can be approximately accounted for through a non-resonant background (see Sec.~\ref{sec:cimsolver}).}
\label{fig:bempp}
\end{figure}

\begin{figure}[t]
\includegraphics[width=\textwidth]{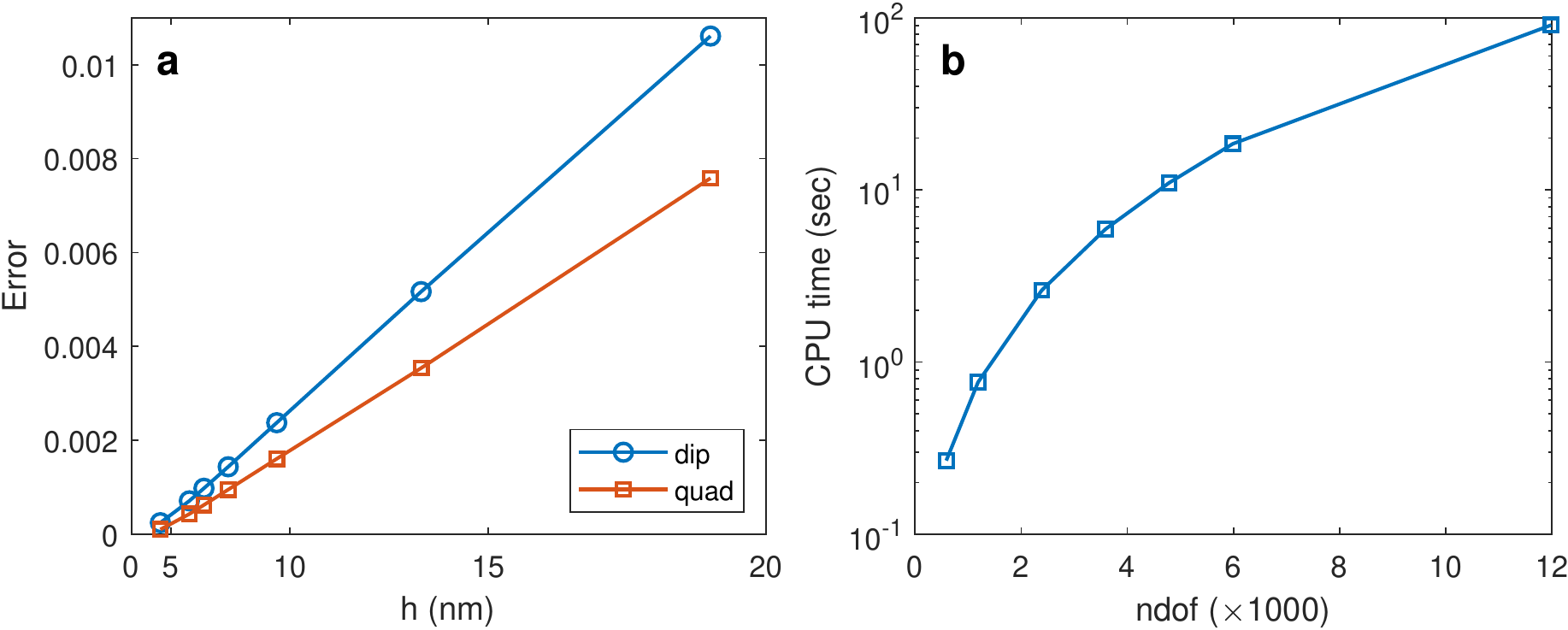}
\caption{Convergence properties and computer time.  (a) Relative error between resonance energies for dipole and quadrupole modes as computed within \cim and Mie theory for the gold nanosphere with a diameter of 100 nm investigated in Fig.~\ref{fig:bempp}(b) and for different boundary discretizations.  $h$ is the mean edge length of the triangulated sphere boundary, for better visibility we use a quadratic scale.  We use sphere discretizations with 100, 200, 400, 600, 800, 1000, and 2000 vertices.  (b) Mean computer times for the solution of the \bem equations at a single frequency for the different sphere discretizations withd an increasing number of global degrees of freedom, as obtained with a standard Intel Core processor i7-7700 CPU at 3.60GHz.  Typical \cim simulations take from less than a minute for coarse discretizations to several hours for fine discretizations.}
\label{fig:convergence}
\end{figure}

The main purpose of this paper is to provide a toolbox for the simple and efficient computation of resonance modes using the \bem approach.  Users interested in such an approach are advised to set up their own problems, and to critically examine them using computer experiments.  From our own experience we found that several things can go wrong in the computation of resonance modes.  In the following we list a few dos and don'ts.

\textit{Accumulation points.}  For nanostructures there usually exist accumulation points where an infinite number of resonance frequencies converges at a single frequency, e.g. the edge modes of a cube.  The \cim algorithm may fail when such an accumulation point falls into the contour $\Gamma$.  In such situations one often finds that all singular values have the same order of magnitude, rather than the standard situation where the large singular values are separated from the smaller ones through a noticeable gap (see Fig.~\ref{fig:cimgui}).

\textit{Accuracy for boundary integration.}  In particular for boundary discretizations where the boundary elements have different sizes, it is important to select high accuracy for triangle integrations.  When the results of \bem and \cim simulations differ, it might be a good idea to increase the accuracy for boundary integrations, see table~\ref{tab:bemoptions} for the control of the quadrature performance using an option structure.

\textit{Accuracy of toolbox.}  We have tried our best to make the implementation of the \bem approach within the \nanobem toolbox as transparent and accurate as possible, where most of the quadrature rules can be directly controlled by the user.  Comparison of our toolbox results with Mie theory shows excellent agreement throughout, demo programs can be obtained in the help pages.  Fig.~\ref{fig:bempp} shows for selected geometries a comparison of our results with those obtained with the independent \textsc{bem}++ software~\cite{bempp:15}, giving excellent agreement throughout.  

\textit{Analytic continuation.}  While a Drude dielectric function of the form
\begin{equation}\label{eq:drude}
  \varepsilon(\omega)=1-\frac{\omega_p^2}{\omega(\omega+i\gamma)}
\end{equation}
can be easily extended to the complex frequency plane, a function tabulated for real frequencies (e.g. \verb!epstable!) cannot.  This doesn't necessarily mean that there will be an error message in the computation of the resonance modes, but the results will be meaningless.  For this reason, users must assure that their permittivity and permeability functions are suitable for a continuation into the complex plane.  For instance, both functions should be causal and thus fulfill the Kramers-Kronig relation.

\textit{BEM and CIM results differ.}  The resonance mode expansion via a finite set of modes is only approximate, and it may happen that the \bem and \cim results differ systematically, for example if not all relevant modes are considered in the \cim.  Also for larger structures a systematic discrepancy between \bem and \cim might show up, see for instance panels (b,d) in Fig.~\ref{fig:bempp}.  See also the discussion about a non-resonant background given below.

\textit{Branch cuts.}  The contour $\Gamma$ of Eq.~\eqref{eq:cimwork1} should avoid crossing branch cuts of the square root of the permittivity and permeability functions.  For Drude dielectric functions similar to Eq.~\eqref{eq:drude} the toolbox has implemented a suitable \verb!zsqrt! function that avoids crossing branch cuts, for other material functions one should be at least aware of this problem.

\textit{Convergence.}  Quite generally, the \bem results should converge to the exact solutions of the transmission problem of Eq.~\eqref{eq:intro1} for sufficiently small mesh sizes $h$ and for sufficiently regular solutions.  Beyond a certain threshold value, which is hard to determine for systems where no analytic solutions are available, the error due to the boundary discretization scales with $h^{\nicefrac 32}$.  As a rule of thumb, the ratio between $h$ and the effective wavelength in the medium should be at least of the order of $1\!:\!5$.  In any case, convergence is a critical issue and should be considered carefully in all simulations.  Fig.~\ref{fig:convergence} shows the convergence properties for selected resonance modes of a sphere towards the exact Mie results, as well as typical simulation runtimes.

\textit{Contour.}  We recommend to always use an elliptic contour for the \cim approach, as it can be shown to give numerically accurate results.

\textit{Cutoff for singular values.}  Singular values with a small magnitude might be due to resonances that lie outside the contour, resonances that lie inside the contour but cannot be resolved well enough due to a too small number of quadrature points along the contour, and numerical artifacts.  For a viable choice of the cutoff parameter, the results of the reconstruction problem should not not depend decisevely on its precise value because modes with a small singular value contribute only little to the expansion of Eq.~\eqref{eq:cim}.  For this reason, the choice of this parameter is usually less critical than one could expect.

\textit{Non-resonant background.}  For larger nanoparticles one occasionally observes a consistent difference between the \bem and \cim results, which can be attributed to a non-resonant background whose origin is not fully understood~\cite{zschiedrich:18,weiss:18,colom:18}.  For a brief discussion and a possibility to account for such a background within the \nanobem toolbox see Sec.~\ref{sec:cimsolver}.

\textit{Number of resonance modes.}  For the optical spectra it usually suffices to keep only a small number of modes in the simulation.  For excitations of oscillating dipoles, see e.g. \texttt{democim02.m}, the number of simulation modes should be larger, in particular for dipoles situated close to the nanoparticle.

\textit{Residuals.}  To check the accuracy of the \cim approach, it might be a good idea to inspect the residuals for the resonance modes.  See Sec.~\ref{sec:cimsolver} for details.

\textit{Resonance frequency tracking.}  When \bem and \cim simulations agree for small structures but differ for larger structures, it might be a good idea to track the modes when continuously increasing the system size.  It can happen that specific resonances move out of the contour in the complex plane and are no longer detected by the \cim algorithm.  In such cases, the contour should be enlarged to encompass all relevant resonance frequencies.

\textit{Spectra versus nearfields.}  It often happens that the optical spectra computed with \bem and \cim are very similar, but the optical nearfields differ.  This is no surprise because the spectra are only governed by the farfields features of the resonance modes.  For this reason, it may be beneficial to compute the scattering spectra from the optical far fields, see for instance \verb!farscattering! command in \texttt{democim01.m}.

\section*{Acknowledgment}

We thank Marko \v{S}imi{\'c} for valuable feedback on the toolbox and Thomas Weiss for a careful reading of the manuscript and most helpful suggestions.  This work has been supported in part by the Austrian Science Fund FWF under project P 31264  and by NAWI Graz.

\appendix

\section{Representation formula}\label{sec:representation}

In this appendix we derive the representation formula~\eqref{eq:representation} and its implementation within the boundary element method approach.  The derivation closely follows the original work of Stratton and Chu~\cite{stratton:39}, which is discussed in more detail in~\cite{chew:95,hohenester:20}, and is repeated here for the sake of completeness.  Our starting point is the wave equation for the electric field
\begin{equation}\label{eq:rep1}
  -\nabla\times\nabla\times\bm E(\bm r)+k^2\bm E(\bm r)=-i\mu\omega\bm J(\bm r)\,,
\end{equation}
where $k$ is the wavenumber inside a given medium and $\bm J(\bm r)$ is an external current distribution.  It turns out to be convenient introducing a Green's function defined through
\begin{equation}\label{eq:rep2}
  -\nabla\times\nabla\times\dy G(\bm r,\bm r')+  k^2\dy G(\bm r,\bm r')=
  -\delta(\bm r-\bm r')\dy I,,
\end{equation}
which gives the electromagnetic response to a unit source located at position $\bm r'$ subject to the boundary condition of outgoing waves at infinity (Silver-M\"uller radiation condition).  It can be shown that the Green's function has the form of an outgoing spherical wave~\cite{chew:95,hohenester:20}
\begin{equation}\label{eq:rep3}
  \dy G(\bm r,\bm r')=\left(\dy I+\frac{\nabla\nabla}{k^2}\right)g(\bm r,\bm r')\,,\qquad
  g(\bm r,\bm r')=\frac{e^{ik|\bm r-\bm r'|}}{4\pi|\bm r-\bm r'|}\,,
\end{equation}
where we have used the dyadic product $(\nabla\nabla)_{ij}=\partial_i\partial_j$ and $\dy I$ is the unit matrix.  The matrix $\dy G$ is often referred to as the dyadic Green's function or Green's dyadics, in short.  Apart from a prefactor, $G_{ij}(\bm r,\bm r')$ relates a current source at position $\bm r'$ oriented along the Cartesian direction $j$ to the $i$'th component of the electric field at position $\bm r$.   We next multiply the wave equation~\eqref{eq:rep1} for the electric field from the right-hand side with the Green's dyadics
\begin{displaymath}
  \Big[-\nabla'\times\nabla'\times\bm E(\bm r')\Big]\cdot\dy G(\bm r',\bm r)+
  k^2\,\bm E(\bm r')\cdot\dy G(\bm r',\bm r)=-
  i\omega\mu\bm J(\bm r')\cdot\dy G(\bm r',\bm r)\,.
\end{displaymath}
For later convenience we have written the wave equation for the primed positions $\bm r'$ rather than the unprimed ones.  Similarly, we multiply Eq.~\eqref{eq:rep2} from the left-hand side with $\bm E(\bm r')$,
\begin{displaymath}
  -\bm E(\bm r')\cdot\nabla'\times\nabla'\times\dy G(\bm r',\bm r)+
  k^2\,\bm E(\bm r')\cdot\dy G(\bm r',\bm r)=-
  \bm E(\bm r')\delta(\bm r'-\bm r)\,.
\end{displaymath}
We next integrate the two equations over $\bm r'$, which is assumed to be located in volume $\Omega$, and subtract them to arrive at~\cite[Eq.~(8.1.18)]{chew:95}
\begin{equation}\label{eq:rep4}
  \int_\Omega\left\{
  \Big[-\nabla'\times\nabla'\times\bm E(\bm r')\Big]\cdot\dy G(\bm r',\bm r)
  +\bm E(\bm r')\cdot\nabla'\times\nabla'\times\dy G(\bm r',\bm r)\right\}
  d^3r'=-\bm E_{\rm inc}(\bm r)+\bm E(\bm r)\,,
\end{equation}
with the incoming electric field
\begin{equation}
  \bm E_{\rm inc}(\bm r)=i\mu\omega\int_{\Omega}\dy G(\bm r,\bm r')\cdot
  \bm J(\bm r')\,d^3r'
\end{equation}
that would be the solution of Maxwell's equations for an unbounded medium in absence of the boundary $\partial\Omega$.  The term in curly brackets of Eq.~\eqref{eq:rep4} can be rewritten as~\cite[Eq.~(8.1.19)]{chew:95}
\begin{equation}\label{eq:rep5}
  -\nabla'\cdot\Bigl\{[\nabla'\times\bm E(\bm r')]\times\dy G(\bm r',\bm r)+
  \bm E(\bm r')\times[\nabla'\times\dy G(\bm r',\bm r)]\Bigr\}\,.
\end{equation}
We next use this modified expression together with Gauss' theorem to obtain a boundary integral of the form
\begin{equation}\label{eq:rep5}
  \bm E(\bm r)=\bm E_{\rm inc}(\bm r)-\oint_{\partial\Omega}
  \Bigl\{[\nabla'\times\bm E(\bm r')]\times\dy G(\bm r',\bm r)+
  \bm E(\bm r')\times\bigl[\nabla'\times\dy G(\bm r',\bm r)\bigr]\Bigr\}
  \cdot\hat{\bm n}'dS'\,,
\end{equation}
where we have explicitly indicated the surface normal $\hat{\bm n}'$ of the boundary.  The first term in curly brackets can be simplified by performing a cyclic permutation of the triple product and using Faraday's law for the curl of $\bm E$, and the second one by performing a cyclic permutation of the triple product.  This leads us to our final expression~\cite[Eq.~(5.26)]{hohenester:20}
\begin{equation}\label{eq:rep}
  \bm E(\bm r)=\bm E_{\rm inc}(\bm r)=-\oint_{\partial\Omega}\left\{i\omega\mu\,\dy G(\bm r,\bm s')
  \cdot\hat{\bm n}'\times\bm H(\bm s')-
  \left[\nabla'\times\dy G(\bm r,\bm s')\right]\cdot\hat{\bm n}'\times\bm E(\bm s')\right\}\,dS'\,.
\end{equation}
In deriving this expression we have exploited that $\dy G$ is a symmetric tensor.  We next introduce the single layer potential~\cite[Eq.~(5.34)]{hohenester:20}
\begin{equation}\label{eq:single}
  \bigl[\mathbb{S}\,\bm u\bigr](\bm r)=\oint_{\partial\Omega}
  \dy G(\bm r,\bm s')\cdot\bm u(\bm s')\,dS'=\oint_{\partial\Omega}\left[
  g(\bm r,\bm s')\bm u(\bm s')+\frac 1{k^2}\nabla g(\bm r,\bm s')\,
  \nabla'\cdot\bm u(\bm s')\right]\,dS'\,,
\end{equation}
where we have used integration by parts to arrive at the second expression, together with the double layer potential
\begin{equation}\label{eq:double}
  \bigl[\mathbb{D}\,\bm u\bigr](\bm r)=\oint_{\partial\Omega}
  \nabla'\times\dy G(\bm r,\bm s')\cdot\bm u(\bm s')\,dS'=-
  \oint_{\partial\Omega}\nabla\times g(\bm r,\bm s')\bm u(\bm s')\,dS'\,
\end{equation}
With these potentials we can cast Eq.~\eqref{eq:rep} to the representation formula~\eqref{eq:representation} given in the main text.  The expressions outside of volume $\Omega$ can be obtained in complete analogy to the above derivation but for the integration over volume $\mathbb{R}^3\backslash\Omega$ and by observing that the outgoing fields at infinity go to zero.

In the numerical implementation for the \bem approach we have to evaluate the single and double layer potentials for the Raviart-Thomas shape functions, see Eq.~\eqref{eq:singlebem}, to arrive at the expressions
\begin{subequations}\label{eq:bempotential}
\begin{eqnarray}
  \mmatrix[\big]{\mathbb S}_{ia,i'a'}
  &=& \oint_{\tau_i}\oint_{\tau_{i'}} \left[
  \bm f_{ia}^e(\bm s)\cdot\bm f_{i'a'}^e(\bm s')-
  \frac{\nabla\cdot\bm f_{ia}^e(\bm s)\,\nabla'\cdot\bm f_{i'a'}^e(\bm s')}{k^2}\right]g(\bm s,\bm s')\,dSdS'\qquad\\
  \mmatrix[\big]{\mathbb D}_{ia,i'a'}&=&\oint_{\tau_i}\oint_{\tau_{i'}}
  \bm f_{ia}^e(\bm s)\cdot\nabla' g(\bm s,\bm s')\times\bm f_{i'a'}^e(\bm s')\,dSdS'\,,
\end{eqnarray}
\end{subequations}
where we have performed for $\mathbb{S}$ an integration by parts in order to bring the derivative from the Green's function to the Raviart-Thomas shape element.  The potentials of Eq.~\eqref{eq:bempotential} are at the heart of our \bem approach.  More details about the numerical evaluation will be presented in \ref{sec:details}.

\section{Details about internal toolbox functions}\label{sec:details}

\begin{table}
\caption{List of the internal classes of the \nanobem toolbox.  The \texttt{tensor} class is described in \ref{sec:detailstensor}, the quadrature classes in \ref{sec:quaddetails}, and the quadrature engine in \ref{sec:bemdetails}.}\label{tab:nanobemdetails}
{\footnotesize
\begin{tabularx}{\columnwidth}{lX}
\hline\hline
Class name & Description \\
\hline
\texttt{tensor} & Tensor class for simple manipulation of multidimensional arrays \\
\texttt{quadboundary} & Quadrature rules for integration over boundary elemens \\
\texttt{quadduffy} & Specialized quadrature rules for touching triangle pairs ~\cite{taylor:03,sarraf:14} \\
\texttt{potbase3} & Auxiliary class for analytic integration over triangle pairs~\cite{haenninen:06} \\[6pt]
\texttt{galerkin.pot1.base} & Base class for evaluation of single and double layer potentials, Eq.~\eqref{eq:bem} \\
\texttt{galerkin.pot1.std} & Standard evaluation over triangle pairs with low accuracy \\
\texttt{galerkin.pot1.refine1} & Refined integration over close triangle pairs w/o series expansion \\
\texttt{galerkin.pot1.refine2} & Refined integration over close triangle pairs with series expansion \\
\texttt{galerkin.pot1.duffy1} & Duffy-type integration over touching triangle pairs w/o series expansion \\
\texttt{galerkin.pot1.duffy2} & Duffy-type integration over touching triangle pairs with series expansion \\
\texttt{galerkin.pot1.smooth1} & Partially analytic integration over triangle pairs w/o series expansion \\
\texttt{galerkin.pot1.smooth2} & Partially analytic integration over triangle pairs with series expansion \\[6pt]
\texttt{galerkin.pot2.base} & Base class for evaluation of representation formula, Eq.~\eqref{eq:representation} \\
\texttt{galerkin.pot2.std} & Standard evaluation over triangle with low accuracy \\
\texttt{galerkin.pot2.smooth} & Partially analytic integration over triangle~\cite{haenninen:06} \\
\hline
\hline
\end{tabularx}}
\end{table}

In this appendix we provide information about the internal toolbox functions and how their performance can be controlled through property pairs.  Table~\ref{tab:nanobemdetails} provides a list for the most important toolbox classes.

\subsection{Tensor class}\label{sec:detailstensor}

The \nanobem toolbox provides a simple tensor class for the evaluation of expressions like
\begin{displaymath}
\quad a_{ij}b_j\,,\quad \sum_j a_{ij}b_j\,,\quad A_{ikj}b_{ij}\,,\dots\,.
\end{displaymath}
The tensor class avoids the time consuming \verb!for! loops of \textsc{matlab} and uses the faster \verb!bsxfun! command instead.  We first introduce dummy expressions for the indices
\begin{code}
[ i, j, k ] = deal( 1, 2, 3 );
\end{code} 
Inside the toolbox functions a value of 1 then represents index $i$, a value of 2 represents index $j$, and so on, with the only important restriction that all indices must have different values.  Tensor objects are initialized via
\begin{code}
a = tensor( rand( 3, 4 ), [ i, j ] );
b = tensor( rand( 4, 1 ), j );
A = tensor( rand( 3, 5, 4 ), [ i, k, j ] );
\end{code}
We here use random arrays for demonstration purposes only.  Once the tensor objects have been initialized, they can be manipulated through
\begin{code}
c1 = a * b;
c2 = sum( a * b, j );
c3 = A * b;
\end{code}
Tensor objects can be manipulated with other tensor objects and scalars, but not with regular \textsc{matlab} arrays.  The usual operators such as \verb!+!, \verb!-!, \verb!*!, \verb!./!, \verb!sin!, \verb!cos!, \verb!exp! have been overloaded for the tensor class and make use of the builtin \verb!bsxfun! command of \textsc{matlab}.  The tensor class is particularly useful for clear and compact coding, when speed is an issue the use of tensor objects should be examined critically.  Note that internally the consistency of the dimensions for the various tensor arrays is not checked.  Finally, conversion of the tensor object back to a normal \textsc{matlab} array is done through
\begin{code}
c1 = double( c1, [ i, j ] );
c2 = double( c2, i );
c3 = double( c3, [ i, j, k ] );
\end{code}
As a representative example we consider the case where the electromagnetic fields for a planewave excitation are evaluated at the positions \verb!pos! on the boundary of the nanoparticle.  The light polarizations \verb!pol! and directions \verb!dir! are arrays of dimension $\mbox{\tt npol}\times 3$, and \verb!pos! is an array of dimension $\mbox{\tt ntau}\times\mbox{\tt nq}\times 3$, where \verb!ntau! and \verb!nq! are the number of boundary elements and quadrature points per element, respectively.  We also provide the material object \verb!mat! for the background medium and the wavenumber \verb!k0!.  The electromagnetic fields can then be computed with the following function:
\begin{code}
function [ e, h ] = fields( pol, dir, mat, k0, pos )
  [ i, q, ipol, k ] = deal( 1, 2, 3, 4 );   %  dummy indices for tensor class
  pol = tensor( pol, [ ipol, k ] );    %  convert polarizations to tensor class
  dir = tensor( dir, [ ipol, k ] );    %  convert directions to tensor class
  pos = tensor( pos, [ i, q, k ] );    %  convert positions to tensor class
  e = exp( 1i * mat.k( k0 ) * dot( pos, dir, k ) ) * pol;
  h = mat.Z( k0 ) * cross( dir, e, k );
  e = double( e, [ i, q, k, ipol ] );  %  convert E to numeric array
  h = double( h, [ i, q, k, ipol ] );  %  convert H to numeric array
end
\end{code}

\subsection{Boundary integration}\label{sec:quaddetails}

At present the toolbox can only deal with triangular boundary elements, but it is written such that in the future also quadrilateral elements might be considered.  For triangle integrations, we first set up a structure
\begin{code}
quad3 = struct( 'x', x, 'y', y, 'w', w );
\end{code}
Here, \verb!x!, \verb!y! are the quadrature points for a unit triangle, and \verb!w! are the integration weights.  A number of triangle quadrature rules is provided through 
\begin{code}
quad3 = triquad( rule );               %  integration rules 1-19
quad3 = triquad( rule, 'refine', 2 );  %  split triangle
\end{code}
Once the integration points and weights for the unit triangle are specified, they can be used for the initialization of the quadrature rules for boundary integration
\begin{code}
pt = quadboundary( tau, 'quad3', quad3 );
\end{code}
\verb!pt! is an \verb!quadboundary! object or an array thereof, with the elements
\begin{code}
pt.tau      %  boundary elements with same materials at inside and outside
pt.quad     %  quadrature rules for boundary elements
pt.npoly    %  number of edges of boundary element, 3 for triangles
pt.mat      %  material vector
pt.inout    %  index [i1,i2] to material vector, i1 for inside and i2 for outside
\end{code}
Each \verb!quadboundary! object always groups boundary elements \verb!tau! with the same shape (currently only triangles) and with the same materials at the inside and outside.  This can considerably simplify the evaluation of boundary integrals, as will be demonstrated below.  With
\begin{code}
[ pos, w, f ] = eval( pt );
[ pos, w, f, fp ] = eval( pt );
\end{code}
we obtain the quadrature positions of dimension $\mbox{\tt ntau}\times \mbox{\tt nq}\times 3$, where \verb!ntau! is the number of boundary elements \verb!pt.tau! and \verb!nq! the number of quadrature points per boundary element, the quadrature weights \verb!w! of dimension $\mbox{\tt ntau}\times \mbox{\tt nq}$, and the Raviart-Thomas shape elements \verb!f! of Eq.~\eqref{eq:shape} with dimension $\mbox{\tt ntau}\times \mbox{\tt nq}\times\mbox{\tt npoly}\times 3$, where \verb!npoly! labels the three shape elements associated with the triangle edges.  The additional output argument \verb!fp! gives the divergence of the shape elements $\nabla\cdot\bm f_\nu^e$ of size $\mbox{\tt ntau}\times \mbox{\tt nq}\times\mbox{\tt npoly}$, which is needed in some cases.

As a representative example, in the following we show how to compute the inhomogeneities of Eq.~\eqref{eq:qinc} for a planewave excitation
\begin{equation}\label{eq:qinc2}
  \mmatrix{q_{E}^{\rm inc}}_{ia}=\oint_{\tau_i}\bm f_{ia}^e(\bm s)\cdot\bm E^{\rm inc}(\bm s)\,dS\,,
\end{equation}
with a similar expression for the magnetic field.  Note that the shape function $\bm f_{ia}^e$  is here defined for a specific triangle $\tau_i$ and for edge $a$, the translation to the global degrees of freedom $\nu$ will be done at the end.  Using the field evaluation function of \ref{sec:detailstensor} and a single \verb!quadboundary! object \verb!pt!, the inhomogeneities can be computed through
\begin{code}
function [ qe, qh ] = qfields( pol, dir, mat, k0, pt )
  [ pos, w, f ] = eval( pt ); 
  [ e, h ] = fields( pol, dir, mat, k0, pos );
  [ i, ipol, q, a, k ] = deal( 1, 2, 3, 4, 5 );
  e = tensor( e, [ i, q, k, ipol ] );
  h = tensor( h, [ i, q, k, ipol ] );
  f = tensor( f, [ i, q, a, k ] );  
  w = tensor( w, [ i, q ] );
  qe = sum( dot( f, e, k ) * w, q );  qe = double( qe, [ i, a, ipol ] );
  qh = sum( dot( f, h, k ) * w, q );  qh = double( qh, [ i, a, ipol ] );
end
\end{code}
We can finally use this function for the total evaluation of the inhomogeneities of Eq.~\eqref{eq:qinc}.  In a first step we allocate the arrays for the inhomogeneities, using the number of global degrees of freedom \verb!ndof(tau)! for the boundary elements.  We then loop over the boundary elements and evaluate the inhomogeneities of Eq.~\eqref{eq:qinc2} provided that the boundary is connected to the embedding medium.  Finally, we translate from the local degrees of freedom $i,a$ to the global degrees~$\nu$.
\begin{code}
[ qinc.e, qinc.h ] = deal( zeros( ndof( tau ), size( pol, 1 ) );  %  allocation
for pt = quadboundary( tau, 'quad3', triquad( 3 ) )  %  loop over boundary elements
  if pt.inout( 2 ) == 1
    [ qe, qh ] = qfields( pol, dir, mat, k0, pt );   %  field evalulation
    qe = reshape( qe, [], size( pol, 1 ) ); 
    qh = reshape( qh, [], size( pol, 1 ) );
    nu = vertcat( pt.tau.nu );  %  translation to global degrees of freedom
    for it = 1 : numel( nu ) 
      qinc.e( nu( it ), : ) = qinc.e( nu( it ), : ) + qe( it, : );
      qinc.h( nu( it ), : ) = qinc.h( nu( it ), : ) + qh( it, : );
    end  
  end
end
\end{code}

\subsection{Quadrature engine}\label{sec:bemdetails}

For the evaluation of the single and double layer potentials we have to evaluate integrals of the form~\cite{hohenester:20}
\begin{equation}\label{eq:divergent}
  \mathcal{I}_{ia,i'a'} =
  \oint_{\tau_i}\oint_{\tau_{i'}}\bm f_{ia}^e(\bm s)\cdot\bm f_{i'a'}^e(\bm s')
  \frac{e^{ik_j|\bm s-\bm s'|}}{4\pi|\bm s-\bm s'|}\,dSdS'\,.
\end{equation}
The integrand diverges for $\bm s\approx\bm s'$ and one has to be careful in the evaluation of the integral for identical or touching triangle pairs, as well as for triangles that are located close to each other.  The evaluation of integrals of the form of Eq.~\eqref{eq:divergent} is performed in the toolbox through a quadrature engine, which allows to use different quadrature rules (coarse or refined) for different triangle pairs.  We also provide the possibility that the exponential in Eq.~\eqref{eq:divergent} is expanded in a Taylor series, such that the time consuming evaluations of the double integrals have to be performed only once.  The order of the Taylor expansion is controlled by the parameter \verb!order! (with a default value of three).   In the following we describe the quadrature engine and the parameters for its control in slightly more detail.

\begin{figure}[t]
\centerline{\includegraphics[width=0.65\textwidth]{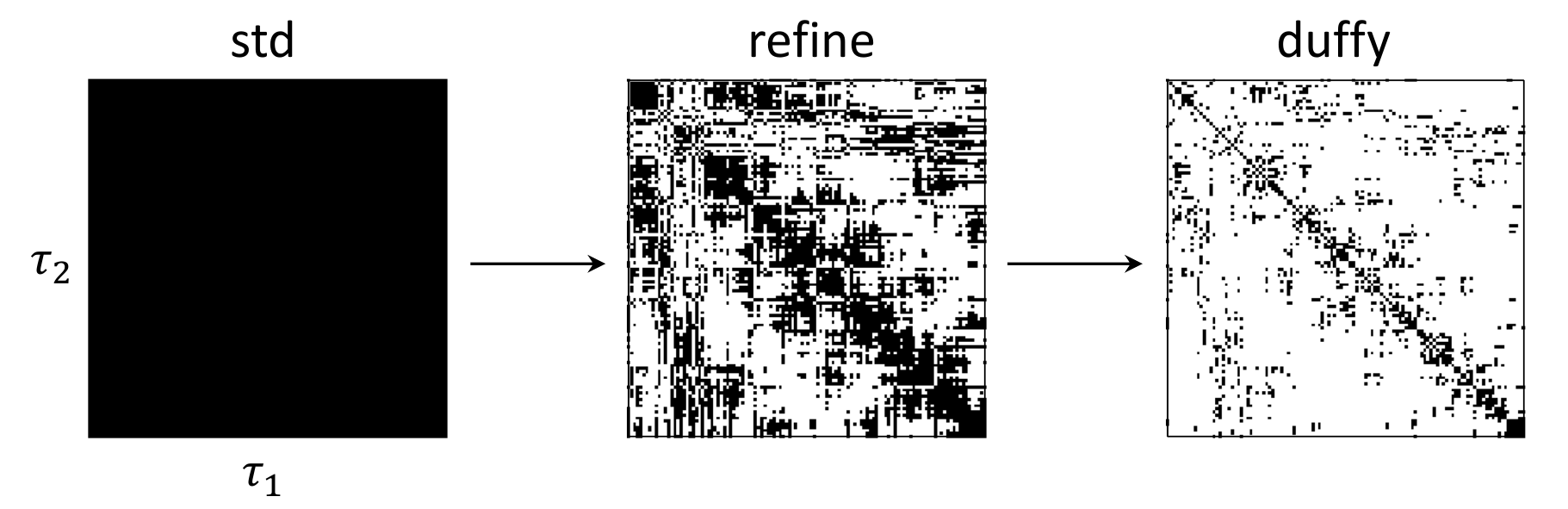}}
\caption{Quadrature engine for the evaluation of integrals over triangle pairs $\mathcal{I}(\tau_1,\tau_2)$.  In the first step \texttt{std}, we evaluate the integrals using quadrature rules of low order with only few quadrature points per triangle.  In the second step \texttt{refine}, we use a refined quadrature rules for triangle pairs that are sufficiently close to each other.  In the final step \texttt{duffy}, we use specialized quadrature rules for identical or touching triangle pairs to deal with the singular integrands of the single and double layer potentials.}\label{fig:quadengine}
\end{figure}

We first introduce a measure for the distance between two triangles $\tau_1$, $\tau_2$.  First, we enclose each boundary element $\tau$ in a sphere with radius $\mbox{rad}(\tau)$ that is centered around $\mbox{pos}(\tau)$.  The mean distance scaled by the bounding box radius is then defined as the relative distance
\begin{equation}
  d(\tau_1,\tau_2)=\frac{\|\mbox{pos}(\tau_1)-\mbox{pos}(\tau_2)\|}{\mbox{rad}(\tau_1)+\mbox{rad}(\tau_2)}\,.
\end{equation}
When this distance is smaller than some parameter \verb!relcutoff! we usually perform a refined integration.  The working principle of our quadrature engine is depicted in Fig.~\ref{fig:quadengine}.  We start with a coarse integration for all triangle pairs, and then successively employ more refined quadrature rules. In each refinement step the previous results are overwritten.  To set up the quadrature engine, we first define the quadrature rules for the standard and refined integration
\begin{code}
rules  = quadboundary.rules( 'quad3', triquad(  3 ) );  %  standard
rules1 = quadboundary.rules( 'quad3', triquad( 11 ) );  %  refined
\end{code}
The quadrature engine is then set up for instance through
\begin{code}
pot( 1 ) = galerkin.pot1.std( 'rules', rules );
pot( 2 ) = galerkin.pot1.refine2( 'relcutoff', 2, 'order', 3, 'rules1', rules1 );
pot( 3 ) = galerkin.pot1.duffy2( 'nduffy', 3 );
\end{code}
Here \verb!order! controls the Taylor series expansion, \verb!relcutoff! controls the triangle distance for refined integration, and \verb!nduffy! controls the quadrature rules for touching or identical triangle pairs.  The quadrature engine can be passed to the \bem solver through
\begin{code}
bem = galerkin.bemsolver( tau, 'engine', pot );
\end{code}

\begin{table}
\caption{The performance of the \texttt{galerkin.bem} and \texttt{galerkin.cim} solvers can be controlled through a number of options that can be passed in the form of property names and values to the solvers.  Alternatively one can also use a option structure, in complete analogy to the \texttt{inputParser} object of\textsc{matlab}.}\label{tab:bemoptions}
{\footnotesize
\begin{tabularx}{\columnwidth}{llX}
\hline\hline
Property name & Default value & Description \\
\hline
\texttt{'relcutoff'} & 2 & Cutoff for refined triangle integrations \\
\texttt{'rules'} & \texttt{triquad(3)} & Default integration rules, see \texttt{quadboundary.rules} \\
\texttt{'rules1'} & \texttt{triquad(11)} & Quadrature rules for refined integration \\
\texttt{'order'} & 3 & Order of series expansion, \texttt{[]} if none \\
\texttt{'nduffy'} & 3 & Parameter controlling Duffy integration rules for touching triangles \\
\texttt{'waitbar'} & 0 & Show waitbar during initialization \\
\hline
\hline
\end{tabularx}}
\end{table}

\noindent In the evaluation of the single and double layer potentials we then progress in \verb!bem! through the \verb!pot! array, where each element provides the quadrature rules for the coarse or refined triangle integrations.  Alternatively, we can also set up an option array and pass the arguments controlling the engine directly to the \bem solver (see also Table~\ref{tab:bemoptions})
\begin{code}
op = struct( 'relcutoff', 2, 'order', 3, ... );
bem = galerkin.bemsolver( tau, op );
\end{code}
Note that with \verb!order=[]! no series expansion is performed in the exponential of Eq.~\eqref{eq:divergent}, which can lead to more accurate results at the expense of a significant increase of simulation time.  A similar quadrature engine approach is used when computing the fields away from the boundary, using the representation formula~\eqref{eq:representation}.  The relative distance between position $\bm r$ and the boundary element $\tau$ is defined through the distance between $\bm r$ and $\mbox{pos}(\tau)$ scaled by the bounding box radius $\mbox{rad}(\tau)$.  The quadrature engine is set up in the same way as before
\begin{code}
pot( 1 ) = galerkin.pot2.std( 'rules', rules );
pot( 2 ) = galerkin.pot2.smooth( 'relcutoff', 2, 'rules', rules1 );
\end{code}
This engine is used for instance in the \verb!fields! function for the \bem solution vector, where the performance can be controlled again by passing arguments to the function.  The \verb!smooth! integration function uses the singularity subtraction method of~\cite{haenninen:06} for the evaluation of the integrals with divergent integrands.  More details about the performance and functionality of the engine functions can be found in the help pages of the toolbox.



\bibliographystyle{elsarticle-num}
\biboptions{sort&compress}

\begin{thebibliography}{10}
\expandafter\ifx\csname url\endcsname\relax
  \def\url#1{\texttt{#1}}\fi
\expandafter\ifx\csname urlprefix\endcsname\relax\def\urlprefix{URL }\fi
\expandafter\ifx\csname href\endcsname\relax
  \def\href#1#2{#2} \def\path#1{#1}\fi

\bibitem{leung:94}
P.~T. Leung, S.~Y. Liu, K.~Young, Completeness and orthogonality of quasinormal
  modes in leaky optical cavities, Phys. Rev. A 49 (1994) 3057.

\bibitem{kristensen:12}
P.~T. Kristensen, C.~V{an V}lack, S.~Hughes, Generalized effective mode volume
  for leaky optical cavities, Opt. Lett. 37 (2012) 1649.

\bibitem{bai:13}
Q.~Bai, M.~Perrin, C.~Sauvan, J.-P. Hugonin, P.~Lalanne, Efficient and
  intuitive method for the analysis of light scattering by a resonant
  nanostructure, Opt. Express 21~(22) (2013) 27371--27382.

\bibitem{sauvan:13}
C.~Sauvan, J.~P. Hugonin, I.~S. Maksymov, P.~Lalanne, Theory of the spontaneous
  optical emission of nanosize photonic and plasmon resonators, Phys. Rev.
  Lett. 110 (2013) 237401.

\bibitem{ge:14}
R.~C. Ge, P.~T. Kristensen, J.~F. Young, S.~Hughes, Quasinormal mode approach
  to modelling light-emission and propagation in nanoplasmonics, New J. Phys.
  16 (2014) 113048.

\bibitem{doost:14}
M.~B. Doost, W.~Langbein, E.~A. Muljarov, Resonant-state expansion applied to
  three-dimensional open optical systems, Phys. Rev. A 90 (2014) 013834.

\bibitem{muljarov:16}
E.~A. Muljarov, W.~Langbein, Resonant-state expansion of dispersive open
  optical systems: Creating gold from sand, Phys. Rev. B 93 (2016) 075417.

\bibitem{perrin:16}
M.~Perrin,
  \href{http://www.opticsexpress.org/abstract.cfm?URI=oe-24-24-27137}{Eigen-energy
  effects and non-orthogonality in the quasi-normal mode expansion of Maxwell
  equations}, Opt. Express 24~(24) (2016) 27137--27151.

\bibitem{lalanne:19}
P.~Lalanne, W.~Yan, A.~Gras, C.~Sauvan, J.-P. Hugonin, M.~Besbes, G.~Demesy,
  M.~D. Truong, B.~Gralak, F.~Zolla, A.~Nicole, F.~Binkowski, L.~Zschiedrich,
  S.~Burger, J.~Zimmerling, R.~Remis, P.~Urbach, H.~T. Liu, T.~Weiss,
  Quasinormal mode solvers for resonators with dispersive materials, J. Opt.
  Soc. Am. A 36 (2019) 686.

\bibitem{kristensen:20}
P.~T. Kristensen, K.~Herrmann, F.~Intravaia, K.~Busch,
  \href{http://www.osapublishing.org/aop/abstract.cfm?URI=aop-12-3-612}{Modeling
  electromagnetic resonators using quasinormal modes}, Adv. Opt. Photon. 12
  (2020) 612--708.

\bibitem{schuller:10}
J.~A. Schuller, E.~S. Barnard, W.~Cai, Y.~C. Jun, J.~S. White, M.~L.
  Brongersma, Plasmonics for extreme light concentration and manipulation,
  Nature Mat. 9 (2010) 193.

\bibitem{hohenester:20}
U.~Hohenester, Nano and Quantum Optics, Springer, 2020.

\bibitem{hesthaven:02}
J.~S. Hesthaven, T.~Warburton, High-order/spectral methods on unstructured
  grids {I}. Time-domain solution of {M}axwell's equations, J. Comput. Phys.
  181 (2002) 186.

\bibitem{hesthaven:03}
J.~S. Hesthaven, High-order accurate methods in time-domain computational
  electromagnetics: A review, Advances in Imaging and Electron Physics 127
  (2003) 59.

\bibitem{chew:95}
W.~C. Chew, Waves and fields in inhomogeneous media, IEEE Press, Picsatoway,
  1995.

\bibitem{kern:09}
A.~M. Kern, O.~J.~F. Martin, Surface integral formulation for 3d simulations of
  plasmonics and high permittivity nanostructures, J. Opt. Soc. Am. A 26 (2009)
  732.

\bibitem{franke:19}
S.~Franke, S.~Hughes, M.~K. Dezfouli, P.~T. Kristensen, K.~Busch, A.~Knorr,
  M.~Richter, Quantization of quasinormal modes for open cavities and plasmonic
  cavity quantum electrodynamics, Phys. Rev. Lett. 122 (2019) 213901.

\bibitem{unger:18}
G.~Unger, A.~Tr{\"ug}ler, U.~Hohenester, Novel modal approximation scheme for
  plasmonic transmission problems, Phys. Rev. Lett. 121 (2018) 24680.

\bibitem{hohenester.cpc:12}
U.~Hohenester, A.~Tr{\"u}gler, {MNPBEM - A Matlab Toolbox for the simulation of
  plasmonic nanoparticles}, Comp. Phys. Commun. 183 (2012) 370.

\bibitem{hohenester.cpc:14b}
U.~Hohenester, Simulating electron energy loss spectoscopy with the {MNPBEM}
  toolbox, Comp. Phys. Commun. 185 (2014) 1177.

\bibitem{hohenester.cpc:18}
U.~Hohenester, {Making simulations with the MNPBEM toolbox big: Hierarchical
  matrices and iterative solvers}, Comp. Phys. Commun. 222 (2018) 209.

\bibitem{garcia:02}
F.~J. Garc{\'i}{a de Abajo}, A.~Howie, Retarded field calculation of electron
  energy loss in inhomogeneous dielectrics, Phys. Rev. B 65 (2002) 115418.
  
\bibitem{sauvan:21}
C.~Sauvan, Quasinormal mode expansion for nanoresonators made of absorbing dielectric
  materials: study of the role of static modes, Opt. Express 29 (2021) 8268.

\bibitem{beyn:12}
W.-J-Beyn, An integral method for solving nonlinear eigenvalue problems, Linear
  Algebra and its Applications 436 (2012) 3839.

\bibitem{chang:77}
Y.~Chang, R.~Harrington, A surface formulation for characteristic modes of
  material bodies, IEEE Transactions on Antennas and Propagation 25~(6) (1977)
  789--795.

\bibitem{poggio:73}
A.~J. Poggio, E.~K. Miller,
  \href{https://www.sciencedirect.com/science/article/pii/B9780080168883500088}{Chapter
  4: Integral equation solutions of three-dimensional scattering problems}, in:
  R.~Mittra (Ed.), Computer Techniques for Electromagnetics, International
  Series of Monographs in Electrical Engineering, Pergamon, 1973, pp. 159 --
  264.

\bibitem{wu:77}
T.~K. Wu, L.~L. Tsai, Scattering from arbitrarily-shaped lossy dielectric
  bodies of revolution, Radio Science 12~(5) (1977) 709--718.

\bibitem{zolla:18}
F.~Zolla, A.~Nicolet, G.~Dem\'{e}sy, Photonics in highly dispersive media: the
  exact modal expansion, Opt. Lett. 43 (2018) 5813--5816.

\bibitem{kozlov:99}
V.~Kozlov, V.~Maz\v{c}ya, {Differential Equations with Operator Coefficients
  with Applications to Boundary Value Problems for Partial Differential
  Equations}, Springer, Berlin, 1999.

\bibitem{mennicken:03}
R.~Mennicken, M.~M{\"o}ller, {Non-{S}elf-{A}djoint {B}oundary {E}igenvalue
  {P}roblems}, North-Holland Publishing Co., Amsterdam, 2003.

\bibitem{bykov:13}
D.~A. Bykov, L.~L. Doskolovich, Numerical methods for calculating poles of the
  scattering matrix with applications in grating theory, Journal of Lightwave
  Technology 31 (2013) 793--801.

\bibitem{zschiedrich:18}
L. Zschiedrich, F. Binkowski, N. Nikolay, O. Benson, G. Kewes, and S. Burger, 
  Riesz-projection-based theory of light-matter interaction in dispersive nanoresonators,
  Phys. Rev. A 98 (2018) 043806.  

\bibitem{weiss:18} 
T. Weiss and E. A. Muljarov, How to calculate the pole expansion of the optical 
  scattering matrix from the resonant states, Phys. Rev. B 98 (2018) 085433. 

\bibitem{colom:18}
R. Colom, R. McPhedran, B. Stout, and N. Bonod, Modal expansion of the scattered 
  field: Causality, nondivergence, and nonresonant contribution, 
  Phys. Rev. B 98 (2018) 085418.

\bibitem{bempp:15}
W.~{\'S}migaj, S.~Arridge, T.~Betcke, J.~Phillips, J.~Schweiger, {Solving
  Boundary Integral Problems with {BEM}++}, ACM Trans. Math. Software 41~(6)
  (2015) 1--40.

\bibitem{stratton:39}
J.~A. Stratton, L.~J. Chu, Diffraction theory of electromagnetic waves, Phys.
  Rev. 56 (1939) 99--107.

\bibitem{taylor:03}
D.~J. Taylor, Accurate and efficient numerical integration of weakly singulars
  integrals in {Galerkin EFIE} solutions, IEEE Trans. Antennas Propag. 51
  (2003) 2543.

\bibitem{sarraf:14}
S.~Sarraf, E.~Lopez, G.~Rio{s R}odriguez, J.~D{'Elia}, Validation of a
  {G}alerkin technique on a boundary integral equation for creeping flow around
  a torus, Comp. Appl. Math. 33 (2014) 63.

\bibitem{haenninen:06}
I.~H{\"a}nninen, M.~Taskinen, J.~Sarvas, Singularity subtraction integral
  formulae for surface integral equations with RWG, rooftop and hybrid basis
  functions, Progress in Electromagnetic Research, PIERS 63 (2006) 243.

\end{thebibliography}







\end{document}